\documentclass[11pt]{article}
\usepackage{amssymb,amsfonts}
\usepackage{graphics,amsmath}
\usepackage{subfigure}
\usepackage{graphicx}
\usepackage{verbatim}
\usepackage{color}
\usepackage{hyperref}
\usepackage{lscape}%mode paysage
\hypersetup{colorlinks}
\newcommand{\xdownarrow}[1]{%
  {\left\downarrow\vbox to #1{}\right.\kern-\nulldelimiterspace}
}
\newcommand{\xupdownarrow}[1]{%
  {\left\updownarrow\vbox to #1{}\right.\kern-\nulldelimiterspace}
}

\newcommand\bleu[1]{\textcolor{blue}{#1}}
\newcommand\rouge[1]{\textcolor{red}{#1}}

\definecolor{darkred}{rgb}{1,0,0}
\definecolor{darkgreen}{rgb}{0,0.5,0}
\definecolor{darkblue}{rgb}{0,0,1}
\definecolor{orange}{rgb}{1,0.5,0}
\definecolor{green}{rgb}{0,1,0}
\definecolor{purple}{rgb}{.5,0,1}

\hypersetup{ colorlinks,
linkcolor=darkred,
filecolor=darkgreen,
urlcolor=darkblue,
citecolor=darkblue,
linktocpage=true }

\numberwithin{equation}{section}
\newtheorem{definition}{Definition}[section]

\newtheorem{proposition}[definition]{Proposition}
\newtheorem{theorem}[definition]{Theorem}

\newtheorem{lemma}[definition]{Lemma}
\newtheorem{remark}[definition]{Remark}

\newcommand{\prf}{\underline{Proof:}\ }
\newcommand{\finprf}{\null \hfill {\rule{5pt}{5pt}}\\ \null}
\newcommand{\ie}{{\it i.e.}\ }
\newcommand{\lda}{{\lambda}}

\definecolor{markcolor2}{rgb}{1,0,0}

\definecolor{markcolor3}{rgb}{0,1,0}

\def\hybrid{\topmargin 0pt    \oddsidemargin 0.1in %%%%%%%%%%%%%% Archive-30pt
        \headheight 0pt \headsep 0pt
        \textwidth 16.0cm      % A4 paper
        \textheight 22,2cm       % A4 paper
        \marginparwidth .875in
        \parskip 5pt plus 1pt   \jot = 1.5ex}

\hybrid

\catcode`\@=11

\def\marginnote#1{}

\def\draftlabel#1{{\@bsphack\if@filesw {\let\thepage\relax
   \xdef\@gtempa{\write\@auxout{\string
      \newlabel{#1}{{\@currentlabel}{\thepage}}}}}\@gtempa
   \if@nobreak \ifvmode\nobreak\fi\fi\fi\@esphack}
        \gdef\@eqnlabel{#1}}
\def\@eqnlabel{}
\def\@vacuum{}
\def\draftmarginnote#1{\marginpar{\raggedright\scriptsize\tt#1}}

\def\draft{\oddsidemargin -.5truein
        \def\@oddfoot{\sl preliminary draft \hfil
        \rm\thepage\hfil\sl\today\quad\militarytime}
        \let\@evenfoot\@oddfoot \overfullrule 3pt
        \let\label=\draftlabel
        \let\marginnote=\draftmarginnote
   \def\@eqnnum{(\theequation)\rlap{\kern\marginparsep\tt\@eqnlabel}%
\global\let\@eqnlabel\@vacuum}  }

\def\draft2{
        \def\@oddfoot{\sl preliminary draft \hfil
        \rm\thepage\hfil\sl\today\quad\militarytime}
        \let\@evenfoot\@oddfoot \overfullrule 3pt
        \let\label=\draftlabel
        \let\marginnote=\draftmarginnote
   \def\@eqnnum{(\theequation)\rlap{\kern\marginparsep\tt\@eqnlabel}%
\global\let\@eqnlabel\@vacuum}  }

\def\preprint{\twocolumn\sloppy\flushbottom\parindent 2em
        \leftmargini 2em\leftmarginv .5em\leftmarginvi .5em
        \oddsidemargin -.5in    \evensidemargin -.5in
        \columnsep .4in \footheight 0pt
        \textwidth 10.in        \topmargin  -.4in
        \headheight 12pt \topskip .4in
        \textheight 6.9in \footskip 0pt
        \def\@oddhead{\thepage\hfil\addtocounter{page}{1}\thepage}
        \let\@evenhead\@oddhead \def\@oddfoot{} \def\@evenfoot{} }

\def\numberbysection{\@addtoreset{equation}{section}
        \def\theequation{\thesection.\arabic{equation}}}

\def\underline#1{\relax\ifmmode\@@underline#1\else
        $\@@underline{\hbox{#1}}$\relax\fi}

\makeatletter
\newcounter{pubctr}
\def\publist{\@ifnextchar[{\@publist}{\@@publist}}
\def\@publist[#1]{\list
        {[\arabic{pubctr}]\hfill}{\settowidth\labelwidth{[999]}
        \leftmargin\labelwidth
        \advance\leftmargin\labelsep
        \@nmbrlisttrue\def\@listctr{pubctr}
        \setcounter{pubctr}{#1}\addtocounter{pubctr}{-1}}}
\def\@@publist{\list
        {[\arabic{pubctr}]\hfill}{\settowidth\labelwidth{[999]}
        \leftmargin\labelwidth
        \advance\leftmargin\labelsep
        \@nmbrlisttrue\def\@listctr{pubctr}}}
 \relax
\makeatother

\def\be{\begin{equation}}
\def\ee{\end{equation}}
\def\bea{\begin{eqnarray}}
\def\eea{\end{eqnarray}}

\newcommand{\ZZ}{\mbox{${\mathbb Z}$}}
\newcommand{\NN}{\mbox{${\mathbb N}$}}
\newcommand{\CC}{\mbox{${\mathbb C}$}}
\newcommand{\1}{\mbox{\hspace{.0em}1\hspace{-.24em}I}}
\def\mato{\left(\begin{matrix}} 
\def\matf{\end{matrix}\right)}

\def\cA{{\cal A}}
\def\cB{{\cal B}}

\def\cI{{\cal I}}
\def\cL{{\cal L}}

\def\cV{{\cal V}}

\begin{document}

\renewcommand{\theequation}{\thesection.\arabic{equation}}
\csname @addtoreset\endcsname{equation}{section}

\newcommand{\eqn}[1]{(\ref{#1})}

\begin{titlepage}
\begin{center}
\strut\hfill

{\Large \bf On the origin of dual Lax pairs and their $r$-matrix structure }

\vskip 0.5in

{\bf Jean Avan$^{a}$, Vincent Caudrelier$^{b}$}
\vskip 0.05in
\vskip 0.02in
\noindent
$^{a}${\footnotesize Laboratoire de Physique Th\'eorique et Mod\'elisation (CNRS UMR 8089),\\
Universit\'e de Cergy-Pontoise, F-95302 Cergy-Pontoise, France}
\\[3mm]
\noindent
$^{b}${\footnotesize  School of Mathematics, University of Leeds,\\ LS2 9JT Leeds, United Kingdom}
\\[3mm]

\end{center}

\vspace{1cm}

\centerline{\bf Abstract}

We establish the algebraic origin of the following observations made previously by the authors and coworkers: (i)
A given integrable PDE in $1+1$ dimensions within the Zakharov-Shabat scheme related to a Lax pair 
can be cast in two distinct, dual Hamiltonian formulations; (ii)
Associated to each formulation is a Poisson bracket and a phase space (which are not compatible in the sense of Magri); (iii) Each matrix in the Lax pair 
satisfies a linear Poisson algebra a la Sklyanin characterized by the 
{\it same} classical $r$ matrix. We develop the general concept of dual Lax pairs and dual Hamiltonian 
formulation of an integrable field theory. We elucidate the origin of the common $r$-matrix structure by tracing it back  
to a single Lie-Poisson bracket on a suitable coadjoint orbit 
of the loop algebra ${\rm sl}(2,\CC) \otimes  \CC (\lambda, \lambda^{-1})$. 
The results are illustrated with the examples of the nonlinear Schr\"odinger and Gerdjikov-Ivanov hierarchies.
\vskip0.5in
{\it This paper is dedicated to the memory of Anjan Kundu}

\vfill

\noindent {\footnotesize {\tt E-mail: Jean.Avan@u-cergy.fr,\ v.caudrelier@leeds.ac.uk }}\\

\end{titlepage}
\vfill \eject

%\tableofcontents

\section{Introduction}

We propose in this paper an elucidation of the mechanisms underlying the appearance of certain Hamiltonian structures in integrable hierarchies of $1+1$-dimensional PDEs. 
We consider integrable PDEs obtained in the well-known Lax pair formulation as a zero-curvature condition for a $2$-dimensional 
connection $\{d/dx - L(\lda), d/dt - M(\lda)\}$, 
realizing the so-called Ablowitz-Kaup-Newell-Segur or AKNS scheme \cite{GGKM,ZS,AKNS}. Our motivation 
stems from our previous papers \cite{C,CK} where the original question of the Hamiltonian formulation of integrable defects in a $1+1$ field theory raised in \cite{VC} led to the construction of 
two Poisson brackets acting on different phase spaces for the same PDE. The two Poisson structures acted respectively on each of the two connection matrices 
$L,M$ in the Lax representation of the PDE, and surprisingly endowed both matrices with a Poisson algebra structure parametrized by {\it the same} classical $r$-matrix. 
This aspect was investigated more systematically in \cite{ACDK}.
This situation differs fundamentally from the Poisson bracket hierarchy a la Magri \cite{magri} where mutually compatible Poisson structures act on the same phase space.

Such a result may be viewed as a coincidence but we prove here that it in fact originates in a systematic construction of $1+1$ dimensional Lax pairs for integrable field theories, from a single  initial $0+1$ dynamical system living on a coadjoint orbit of a loop algebra, suitably identified with the dual of the positive-index half of a loop algebra (in our case based on the Lie algebra $sl(2,\CC)$) as exposed by Flaschka, Newell and Ratiu (FNR) \cite{FNR}. This paper describes the complete consistent construction of the $r$-matrix properties of such Lax pairs. The above mentioned duality of Poisson structures and Lax representations for the same integrable PDE, together with the occurrence of the same $r$ matrix structure in both representations, will thus be accounted for.

Note that the terminology of ``duality'' that we used in \cite{CK,ACDK}  and also in the present paper is not to be confused with 
that of e.g. \cite{Rui,Feh,Feh2}. It is to be understood as a notion similar to the world--volume duality in 
string theory, exchanging as it does the role of the space/time
coordinates of the underlying space time;  whereas dualities such as those found in  \cite{Rui,Feh,Feh2} are target-space dualities exchanging the role of dynamical variables $p,q$.

The structure of the paper follows closely the construction procedure. In Section 2 we recall the basis of the coadjoint orbit construction
\cite{FNR} starting from the canonical matrix $L(\lambda$) defined on a coadjoint orbit of the loop algebra of Laurent series with coefficients in $sl(2, \CC)$
picked to be the negative power series. The complex parameter $\lambda$ defining the evaluation representations of the loop algebra and its expression in formal power series, 
is identified as ``spectral parameter''. 
The standard Adler-Kostant-Symes construction endows the coadjoint orbit 
with two compatible Poisson structures and associated $r$-matrix structures. 
We describe the integrability properties
for the Hamiltonian evolutions triggered by suitable invariant traces, and the associated isospectral evolutions taking the generic form 
$\frac{d}{dt_n} L = [ L^{(n)}, L ]$ where $L^{(n)} \equiv P^+ \lambda^n L$.

We then proceed as follows:

{\bf Step 1} (Section \ref{map_Psi}) {\bf The constraint equations}

Choosing one specific time variable $t_n $ one restricts the dynamical variables in $L$ to obey the $n$-th time evolution. Implementation of this restriction constrains
the dynamical variables in $L$ to obey specific algebraic and differential relations encoded in a map $\Psi_n $. We thus define a restricted phase space, this time however of $t_n$-valued fields, and a projected and $\Psi$-restricted Lax matrix $V_n^{(n)}(\lambda, t_n)\equiv \Psi_n L^{(n)}$. It naturally acquires a canonical ultralocal $r$-matrix structure inherited 
from the $r$-matrix of $L$. 

{\bf Step 2} (section \ref{FNRtoAKNS}) {\bf Integrability after constraint}

We compare:

- The result of applying a given constraint $\Psi_n $ and extending the dynamical variables to being $t_n$-dependent fields, onto every other FNR operator $L^{(k)}$, $k\neq n$. 

- The result of computing the operator describing the $t_k$ time evolution for the $k$-th Hamiltonian deduced from the expansion of the monodromy matrix of the differential operator $\partial_{t_n} - V_n^{(n)}(\lambda , t_n)$ . 

The resulting matrices are identical. Hence the complete hierarchy of commuting FNR flows on the coadjoint orbit restricts consistently under any constraint
procedure of Step 1, to the hierarchy structure of the time Lax operators $L^{(k)}$ deduced from the monodromy of the specific ``space'' Lax operator $L^{(n)}$ picked in Step 1. 

{\bf Step 3} (section \ref{dualHam}) {\bf Duality}

We prove that the zero curvature conditions obtained from the respective two distinct choices of an AKNS scheme in step 1 and 2 \ie either first $t_n$ and then $t_k$ or first $t_k$ and then $t_n$ are identical. As a consequence, the corresponding $1+1$ integrable field theory exhibits two ``dual'' Lax pair representations, characterized by different phase spaces, same $r$-matrix structure, and the exchange of $x=t_n$ and $t=t_k$ as ``space'' vs ``time'' variables in the AKNS formulation. The duality, pointed out and investigated in \cite{ACDK} from a covariant field theory point of view, is therefore identified in a purely algebraic framework as a general feature directly originating in the constraint procedure leading from $0+1$ integrable theories on a coadjoint orbit of infinite dimension to $1+1$ integrable theories. 

The above steps are captured in the diagram below which explains graphically the duality of Hamiltonian formulation of a given integrable PDE. The important and nontrivial results are shown in a box  and correspond to Sections \ref{map_Psi}, \ref{FNRtoAKNS} and \ref{dualHam} respectively. 

For a fixed $n\ge 1$, Step 1 and 2  go all the way down from the FNR picture and its associated $R$-operator formulation 
(the distinction between $r$-matrix and $R$ operator formulation \cite{STS1} will be clear in the text) to the AKNS formulation and its associated $r$-matrix. 
The third step brings in 
the right-hand-side of the diagram, corresponding to a parallel but different choice of initial time in the FNR picture. The commutativity of 
the diagram expresses the notion of dual Hamiltonian formulation of an integrable PDE.

We illustrate our results in Section \ref{examples}, using the NLS equation as a first example to make contact with previous results in \cite{ACDK}.
We also provide an example never treated before from this dual point of view: the Gerdjikov-Ivanov equation \cite{GI}. This is all the more interesting as this 
integrable PDE is not part of the traditional AKNS hierarchy based on $V_1^{(1)}$ but rather, is related to the Kaup-Newell hierarchy \cite{KN}. We end up with some open questions and
connections with other issues in classical integrable systems.

Let us round up this introduction by putting our current results in perspective with respect to the state of the art. 
So far and to the best of our knowledge,
only parts of one of the two sides of this diagram have been used as the basis of the theory of integrable PDEs/classical field theories. 
Although it is impossible to pinpoint faithfully the major results in the huge available literature (see references in e.g. \cite{FT,BBT} for a partial overview) 
on such a simple  linear diagram, let us point out some of the main 
historical developments and how they relate to the present construction. The traditional AKNS picture corresponds to picking e.g. the left part of the diagram, 
say with $n=1$ and $t_1=x$. The literature is then essentially split into two main streams: 

$\bullet$ The top (left) half of the diagram (in blue) in which the classical $r$-matrix approach is not required and not dealt with. 
This part of the literature is formulated in the coadjoint orbit picture \cite{FNR}. 
Most studies on Hamiltonian and multi-Hamiltonian structures involve
formal (differential) algebra based on the loop algebra of ${\rm sl}(2,\CC)$ (and some higher rank generalizations) 
together with the Adler-Kostant-Symes (AKS) construction.  See e.g. Chapter $10$ in \cite{BBT}.

$\bullet$ The bottom (left) part of the diagram (in blue) in which the full power of the classical $r$-matrix to describe Hamiltonian properties of an 
integrable hierarchy
is used. In this approach, the building ingredients are the Lax matrix $L$, i.e. the matrix associated to the ``space part'' of the flat connection in the auxiliary 
problem and its classical $r$-matrix structure. These data allow one to deduce the hierarchy of higher time-shift or Lax-partner matrices from expansion in the spectral parameter of the 
monodromy matrix of the space-shift differential operator $d/dx - L(\lda,x)$. The knowledge of the analytical properties is therefore crucial in this approach (see e.g. \cite{FT}). The higher Lax-partner matrices are by-products of the construction and they are not known to exhibit specific Poisson properties w.r.t to the only Poisson brackets ($r$-matrix for $L$) used in this picture. See e.g. Part I, Chapter $3$ in \cite{FT}.

In addition to this short summary of the research in that area is the paper \cite{R} (see also \cite{KR}). The results 
provide a Hamiltonian description of the hierarchies of integrable equations based on a Lax matrix of the type $V_n^{(n)}$ in our notations, for arbitrary $n\ge 1$. 
The author's starting point is at the AKNS level and he derives the corresponding FNR equation for a fixed but arbitrary $t_n$ (denoted $x$ in the paper). 
So the results could be summarised  
by a set of arrows going from the bottom to the top of the left part of our diagram. 
Concerning the notion of duality that we develop here, and which is absent from \cite{R}, it is worth mentioning a major set of works not easily fitted in our diagram.
It deals with the topic of Hamiltonian properties of stationary manifolds \cite{BN} in an integrable hierarchy. A geometric treatment of this question is 
provided in \cite{R}. In \cite{FH1,FH2}, the idea to swap the roles of the 
two independent variables is used extensively to derive the (multi)-Hamiltonian structure of integrable stationary equations and is traced back to \cite{ABl}. 
The $r$-matrix formalism is not used there and one 
ultimately works with one independent variable, and hence ODEs, by reduction to time-independent fields. 

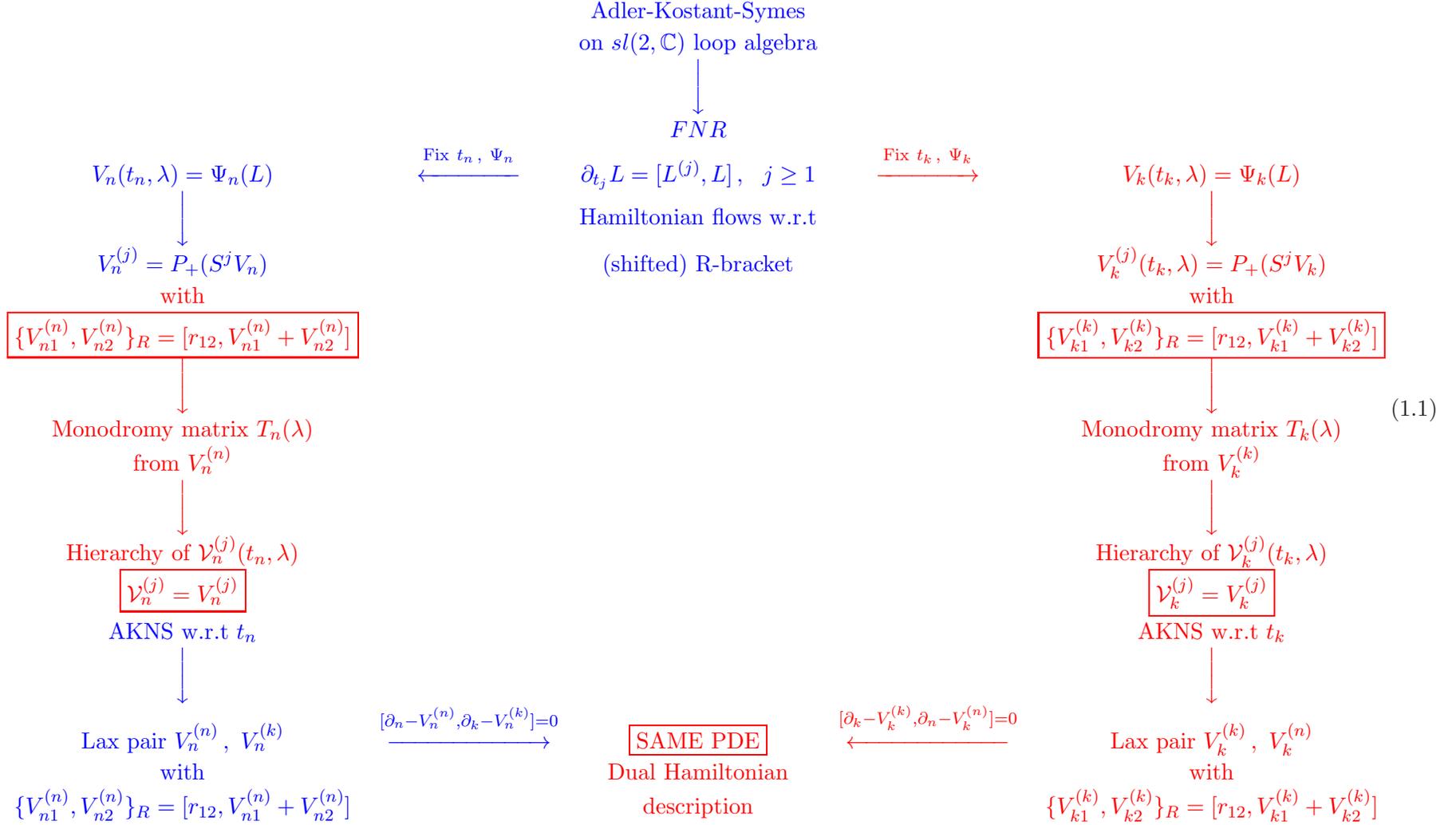
\begin{landscape}
\bea
\begin{array}{ccccc}
& & \bleu{\text{Adler-Kostant-Symes}} & &\\
& & \bleu{\text{on}~ sl(2,\CC)~ \text{loop algebra}} & &\\
 &   & \bleu{\xdownarrow{0.5cm}} &   & \\
& &\bleu{ FNR} & &\\
\bleu{V_n(t_n,\lda)=\Psi_n(L)} & \bleu{\stackrel{\text{Fix}~ t_n\,,~\Psi_n}{\xleftarrow{\hspace*{1.5cm}}}} & \bleu{\partial_{t_j}L=[L^{(j)},L]\,,~~j\ge 1} & \rouge{\stackrel{\text{Fix}~ t_k\,,~\Psi_k}{\xrightarrow{\hspace*{1.5cm}}}} & \rouge{V_k(t_k,\lda)=\Psi_k(L)}\\
  \bleu{\xdownarrow{0.5cm}} &  & \bleu{\text{Hamiltonian flows w.r.t}}  & & \rouge{\xdownarrow{0.5cm}}  \\
\bleu{V_n^{(j)}=P_+(S^j V_n)} & & \bleu{\text{(shifted) R-bracket}}& & \rouge{V_k^{(j)}(t_k,\lda)=P_+(S^j V_k)} \\
\rouge{\text{with}} & & & & \rouge{\text{with}} \\
\rouge{\framebox{$\{V_{n1}^{(n)},V_{n2}^{(n)}\}_R=[r_{12},V_{n1}^{(n)}+V_{n2}^{(n)}]$}} & & & & \rouge{\framebox{$\{V_{k1}^{(k)},V_{k2}^{(k)}\}_R=[r_{12},V_{k1}^{(k)}+V_{k2}^{(k)}]$}} \\
\rouge{\xdownarrow{0.5cm}} &   &  &   & \rouge{\xdownarrow{0.5cm}} \\
 \rouge{\text{Monodromy matrix} ~ T_n(\lda)} &  &  &  & \rouge{\text{Monodromy matrix}~T_k(\lda)}\\
 \rouge{\text{from} ~V_n^{(n)}} &  &  &  & \rouge{\text{from} ~V_k^{(k)}}\\
 \rouge{\xdownarrow{0.5cm}} &   &  &   & \rouge{\xdownarrow{0.5cm}} \\
 \rouge{\text{Hierarchy of}~ \cV_n^{(j)}(t_n,\lda)}& & & & \rouge{\text{Hierarchy of}~ \cV_k^{(j)}(t_k,\lda)}\\
  \rouge{\framebox{$\cV_n^{(j)}=V_n^{(j)}$}}& & & & \rouge{\framebox{$\cV_k^{(j)}=V_k^{(j)}$}}\\
\bleu{  \text{AKNS w.r.t}~ t_n}& & & & \rouge{\text{AKNS w.r.t}~ t_k}\\
\bleu{ \xdownarrow{0.5cm}} &   &  &   & \rouge{\xdownarrow{0.5cm}} \\
  \bleu{\text{Lax pair}~ V_n^{(n)}\,,~ V_n^{(k)}}&\bleu{\stackrel{[\partial_n-V_n^{(n)},\partial_k-V_n^{(k)}]=0}{\xrightarrow{\hspace*{2.5cm}}}} & \rouge{\framebox{SAME PDE}} &\rouge{\stackrel{[\partial_k-V_k^{(k)},\partial_n-V_k^{(n)}]=0}{\xleftarrow{\hspace*{2.5cm}}}} & \rouge{\text{Lax pair}~ V_k^{(k)}\,,~ V_k^{(n)}}\\
\bleu{\text{with}}  &   & \rouge{\text{Dual Hamiltonian}} &   & \rouge{\text{with}}\\
\bleu{\{V_{n1}^{(n)},V_{n2}^{(n)}\}_R=[r_{12},V_{n1}^{(n)}+V_{n2}^{(n)}]} &   & \rouge{\text{description}} &   & \rouge{\{V_{k1}^{(k)},V_{k2}^{(k)}\}_R=[r_{12},V_{k1}^{(k)}+V_{k2}^{(k)}]}\\
  \end{array}
\eea
\end{landscape}

\section{Algebraic construction of the hierarchy: FNR construction}

We first recall the construction of the AKNS hierarchy along the lines of \cite{FNR}. The power of the procedure is to set on equal 
footing all time flows within a well-defined hierarchy from the start. It also provides by construction a Hamiltonian formulation of the hierarchy, 
and the related integrability properties, built in by the application of the Adler-Kostant-Symes procedure \cite{Ad,Ko,Sy} to the ${\rm sl}(2,\CC)$ loop algebra endowed with a second Poisson bracket structure compatible with the canonical one. 

The procedure goes as follows. Let 
$\cL$ be the Lie algebra of Laurent series $X:\lda\mapsto X(\lda)$ with coefficients in ${\rm sl}(2,\CC)$ \ie
\be
X(\lda)=\sum_{j=-\infty}^N X_j \lda^j\,,~~X_j\in{\rm sl}(2,\CC)\,;~~\text{for some integer $N$}\,,
\ee
with the standard Lie bracket
\be
[X,Y](\lda)=\sum_k\sum_{i+j=k}[X_i,Y_j]\lda^k\,.
\ee
There is a natural decomposition of $\cL$ into Lie subalgebras $\cL=\cL_-\oplus \cL_+$ where
\be
\cL_-=\{\sum_{j=-\infty}^{-1} X_j \lda^j\}\,,~~\cL_+=\{\sum_{j=0}^\infty X_j \lda^j\,;~~X_j=0~~\forall j>N~~\text{for some integer $N\ge 0$}\}\,.
\ee
This yields two projectors $P_+$ and $P_-$ and we define the $R$ operator as
\be
\label{R_operator}
R=P_+-P_-\,.
\ee
It is well-known that this operator satisfies the modified classical Yang-Baxter equation and allows one to define a second Lie bracket $[~,~]_R$ on $\cL$ 
(see e.g. \cite{STS_lectures})
\be
\label{R_bracket}
[X,Y]_R=\frac{1}{2}[(RX),Y]+\frac{1}{2}[X,(RY)]\,.
\ee
Using the following ad-invariant nondegenerate symmetric bilinear form on $\cL$, for all $X,Y\in\cL$,
\be
\label{bilinear_form}
\left(X,Y\right)=\sum_{i+j=-1}{\rm Tr}(X_i,Y_j)\equiv {\rm Res}_{\lda}{\rm Tr}(X(\lda)Y(\lda))\,,
\ee
where ${\rm Tr}$ in the first equality is the Killing form on ${\rm sl}(2,\CC)$,
the two Lie brackets allow to define Lie-Poisson brackets on $C^\infty(\cL)$ by setting, for all $F,G \in C^\infty(\cL)$ and $X\in\cL$,
\bea
\label{Lie_Poisson}
&&\{F,G\}(X)=(X,[\nabla F(X),\nabla G(X)])\,,\\
\label{R_bracket2}
&&\{F,G\}_R(X)=(X,[\nabla F(X),\nabla G(X)]_R)\,.
\eea
where $\nabla$ denotes as usual the gradient.

We denote the second Poisson bracket as the $R$-bracket. The main statement of FNR \cite{FNR} is that all equations of the AKNS hierarchy 
can be encapsulated into a single set of mutually commuting Hamiltonian time flows:
\be
\partial_{t_k}Q=[Q^{(k)},Q]\,,~~k=1,2,\dots\,,
\ee
with $\displaystyle Q\in \cL_-^0\equiv \{\sum_{j=-\infty}^{0} X_j \lda^j\}$ and 
\be
\label{Qop}
Q^{(k)}=P_+(S^kQ)\,.
\ee
The shift operator $S^k$ on $\cL$ is defined for all $k\in\ZZ$ by
\be
S^k(X)(\lda)=\lda^kX(\lda)\,.
\ee
The associated mutually Poisson-commuting Hamiltonians turn out to be the dynamical coefficients in the $\lambda$ power expansion of $Tr(L(\lambda)^2$. These objects are indeed identified as:
\be
\label{Casimirs}
\phi_k(L)=-\frac{1}{2}(S^k(L),L)\,,~~k\in\ZZ\,.
\ee
This is a consequence of the Adler-Kostant-Symes scheme applied to this specific splitting once the time-evolution operator is identified by the formula (\ref{Qop}).  
We now recall the main steps leading to the FNR result 
using the systematic reformulation in terms of $R$-operator 
proposed by Semenov-Tian-Shansky \cite{STS1} and adapted to our setting.
\begin{proposition}\cite{STS1}
Let $I$ be a finite or countable set and let $C_n$, $n\in I$ be a collection of Casimir functions for the Lie-Poisson bracket $\{~,~\}$ on $\cL$. Then
\begin{enumerate}
\item The Casimir functions are in involution with respect to the $R$-bracket.

\item The commuting Hamiltonian flows generated by $C_n$ with respect to $\{~,~\}_R$ take the Lax form
\be
\partial_{t_n} L=-\frac{1}{2}[(R\nabla C_n(L)),L]\,,~~\forall L\in\cL\,.
\ee
\end{enumerate}
\end{proposition}
The functions $\phi_k$ in \eqref{Casimirs} provide a family of Casimirs for the Lie-Poisson bracket on $\cL$. Applying the proposition, we obtain
\begin{proposition}
\label{prop_flow1}
For any $k\ge 1$, the equation 
\be
\label{FNR_flows}
\partial_{t_k}Q=[Q^{(k)},Q]
\ee
is the flow generated by the function $\phi_k$ with respect to $\{~,~\}_R$, restricted to $\cL_-^0$.
\end{proposition}
\prf
Note that, consistently with (\ref{Casimirs}) 
\be
\nabla \phi_k(X)=-S^k(X)\,,~~\forall X\in\cL\,.
\ee
Let $F\in C^\infty(\cL)$ then
\be
\label{dynF}
\partial_{t_k}F(X)=\{\phi_k,F\}_R(X)=-\frac{1}{2}\left(X,[(R(S^k(X)),\nabla F(X)]\right)\,.
\ee
Note that $[S^k(X),X]=0$, and $\partial_{t_k}F(X)=(\partial_{t_k}X,\nabla F(X))$ hence we deduce
\be
\label{full_eq}
\partial_{t_k}X=[P_+(S^k(X)),X]\,.
\ee
It remains to show that this equation on $\cL$ can be restricted to $\cL_-^0$. For $X\in\cL$, denote $X=X_++X_-$ as the unique decomposition of $X$  
where\footnote{We draw the reader's attention to the fact that $X_\pm\neq P_\pm(X)$ as the decomposition is performed along 
different subalgebras of $\cL$.} $X_-\in\cL_-^0$. Then \eqref{full_eq} is equivalent to
\be
\label{proj_eq}
\partial_{t_k}(X_++X_-)=-[P_-(S^k(X_-)),X_+]-[P_-(S^k(X_-)),X_-]\,,
\ee
which in turn implies
\be
\partial_{t_k}X_+=-[P_-(S^k(X_-)),X_+]_+\,.
\ee
We can thus consistently set $X_+=0$. Inserting back in \eqref{proj_eq}, we obtain the reduced equation to $\cL_-^0$ as
\be
\partial_{t_k}X_-=-[P_-(S^k(X_-)),X_-]=[P_+(S^k(X_-)),X_-]\,.
\ee
yielding \eqref{FNR_flows} with $Q=X_-$.
\finprf

It is important to note that \eqref{FNR_flows} is a {\bf nonlinear} equation in $Q$. It is different in substance 
from the starting point in strictly-speaking AKNS-type integrable hierarchies consisting of defining resolvents $R$
associated to an operator $\partial_{t_k}-V^{(k)}$ where $V^{(k)}$ is a {\bf given} polynomial of degree $k$ in $\lda$ with coefficients 
in a given Lie algebra (${\rm sl}(2,\CC)$ for the AKNS hierarchy), \ie,
\be
\label{resolvent}
[\partial_{t_k}-V^{(k)},R]=0\,.
\ee
This is the approach developed to a great level of generality for instance in \cite{Dickey_book} (using a formal algebraic formulation). 
Eq \eqref{resolvent} is {\bf linear} in $R$. This allows one 
to develop a full theory of resolvents using the convenient notion of (formal) dressing. It actually lies at the core of the dressing transformation procedure
for Zakharov-Shabat type Lax representations, such as described in e.g. \cite{AB}.

Our purpose now is to make contact with this level of the theory that we shall denote as AKNS formalism from now on, to distinguish it from the FNR formalism discussed so far. 
In a nutshell, we want to transfer the Hamiltonian and $R$-operator content of FNR down to AKNS. To some extent, the Hamiltonian content
is already addressed in \cite{FNR}. We now translate the language of the $R$ operator used so far into the more well-known classical $r$-matrix formalism that is widely used at 
the AKNS level of the theory. We first need to obtain the central FNR equation \eqref{FNR_flows} in a slightly more general manner. We use the notion of {\it intertwining operator}
as defined in \cite{RSTS}. 
 \begin{definition}
 A linear operator $A:\cL\to\cL$ is called intertwining if
 \be
A[X,Y]=[AX,Y]=[X,AY]\,,~~\forall X,Y\in\cL\,.
 \ee
 \end{definition}
\begin{proposition}\cite{RSTS}\\
If $R$ is a solution of the modified classical Yang-Baxter equation and $A$ is an intertwining operator then $\widetilde{R}=R\circ A$ is also a solution of the  modified classical Yang-Baxter equation.
\end{proposition}
As a consequence, one can consistently define a Lie bracket $[~,~]_{\widetilde{R}}$ on $\cL$ and hence the corresponding $\widetilde{R}$-bracket. It is easy to check that the shift operator $S^k$ is an intertwining operator on $\cL$ for all $k\in\ZZ$. 

\begin{definition}
\label{kbra}
The family of Poisson brackets on $\cL$ associated to $R\circ S^k$, $k\in\ZZ$ is denoted by $\{~,~\}_k$.
\end{definition}

We can now revisit Proposition \ref{prop_flow1} and state
\begin{proposition}
\label{prop_flow2}
Let $k\ge 1$ be fixed. The equation 
\be
\label{FNR_flows_k}
\partial_{t_k}Q=[Q^{(k)},Q]
\ee
is the flow generated by the function $\phi_m$ in \eqref{Casimirs} with respect to $\{~,~\}_{k-m}$ for all $m\in\ZZ$, restricted to $\cL_-^0$.
\end{proposition}
\prf
It suffices to note that for all $m,n\in\ZZ$, using \eqref{dynF}, 
\be
\{\phi_m,F\}_n(L)=\{\phi_{m+n},F\}_R(L)\,,~~\forall L\in\cL\,,~~\forall F\in C^\infty(\cL)\,.
\ee
\finprf
We have the following useful lemma
\begin{lemma}
\label{lemma_shift}
Let $k\in\ZZ$,
\be
\{F\circ S^k,G\circ S^k\}_{-k}(L)=\{F,G\}_R(S^kL)\,,~~\forall L\in\cL\,,~~\forall F,G\in C^\infty(\cL)\,.
\ee
\end{lemma}

Now we make the connection with the classical $r$-matrix formalism. It will be convenient to use the so-called auxiliary space notation in the rest of the paper. For instance, for any $X\in\cL$, we write\footnote{Note that we are rather loose in the definition of the object $\1$ 
is not properly defined. Doing so would make the paper even longer. We prefer to take the view that this is a well-established notation for an appropriate identity map which should not lead to any confusion.}
\be
\label{aux_space}
X_1=X\otimes \1\,,~~X_2=\1\otimes X\,.
\ee

Let us extend the bilinear form \eqref{bilinear_form} 
to the tensor product $\cL\otimes \cL$ by setting 
 \be
 (X\otimes Y,Z\otimes W)=(X,Z)(Y,W)\equiv {\rm Res}_{\lda,\mu}{\rm Tr}(X(\lda)Z(\lda)){\rm Tr}(Y(\mu)W(\mu))\,,
 \ee
 for all $X,Y,Z,W\in\cL$.
 Another notation we will use to remember the dependence on $\lda,\mu$ is 
\be
{\rm Res}_{\lda,\mu}(X(\lda)\otimes Y(\mu),Z(\lda)\otimes W(\mu))\,.
\ee
 
 \begin{definition}
 \label{def_r_matrix}
 The classical $r$-matrix is the element of $\cL\otimes \cL$ defined by (again with $X,Y$ any two elements of $\cL$) :
 \be
 (RX,Y)=(r,Y\otimes X)={\rm Res}_{\lda,\mu}(r(\lda,\mu),Y(\lda)\otimes X(\mu))\,,~~\forall X,Y\in\cL\,.
 \ee
 \end{definition}
 
\begin{definition}
\label{def_tensor_form}
Let $L\in\cL$ and $k\in\ZZ$.
The element of $\cL\otimes \cL$ denoted by $\{L_1(\lda),L_2(\mu)\}_k$ is defined by
 \be
 \label{TPB}
{\rm Res}_{\lda,\mu}\left(\{L_1(\lda),L_2(\mu)\}_k,Y(\lda)\otimes X(\mu)\right)=(L,[X,Y]_k)\,,~~\forall X,Y\in\cL\,.
 \ee
 \end{definition}
 
 Here and in the following discussions the notation $X_i$ denotes the element of the (possibly multiple but at least double) 
 tensor power of $\cL$ where $X$ is positioned as the $i$-th factor and all other tensorial factors are taken to be $1$.
 
 With these two definitions, we make contact between the $k$-bracket defined in (\ref{kbra}) and the celebrated Sklyanin formula \cite{Skl}. 
 \begin{proposition}
 \label{Sklyanin_formula}
 The following formula holds
 \be
 \{L_1(\lda),L_2(\mu)\}_k=-\frac{1}{2}[r_k(\lda,\mu),L(\lda)\otimes \1]+\frac{1}{2}[\Pi r_k(\mu,\lda),\1\otimes L(\mu)]\,,
 \ee
 where $\Pi$ is the permutation operator on two copies of $\cL$: $\Pi(X\otimes Y)=Y\otimes X$ and 
\be
r_k(\lda,\mu)=-2\frac{\mu^k}{\lda-\mu}t\,,~~t=\Pi^{sl(2,C)}  ~~\text{(Casimir)}\,.
\ee
In particular, 
 \be
\{L_1(\lda),L_2(\mu)\}_R= \{L_1(\lda),L_2(\mu)\}_0=[\frac{\Pi}{\lda-\mu},L_1(\lda)+L_2(\mu)] ~~\text{(Sklyanin formula)}\,.
 \ee
 \end{proposition}
 It is useful here to make this form of Poisson brackets more explicit.
 Using a basis $\{e^a\lda^i\}$ for $\cL$ where $\{e^a\}$ is a basis for ${\rm sl}(2,\CC)$, write 
\be
\{L_1(\lda),L_2(\mu)\}_k=\left(\{L_1(\lda),L_2(\mu)\}_k\right)_{ij}^{ab}e^a\lda^i\otimes e^b\mu^j ~~\text{(summation implied over repeated indices)}\,,
\ee
and let us use the convenient notation 
 \be
 \{L_i^a,L_j^b\}_k\equiv \left(\{L_1(\lda),L_2(\mu)\}_k\right)_{ij}^{ab}\,.
 \ee
 The following component form of the $k$-bracket is useful
 \be
 \label{bracket_components}
 \{L_i^a,L_j^b\}_k=\epsilon_{ij}^k(K^{-1})_{ac}\,C^{cd}_b\,L^d_{i+j+1-k}\,,
 \ee
 where $C^{cd}_f$ are the structure constants of ${\rm sl}(2,\CC)$, $K$ is the matrix with entry $K_{ab}={\rm Tr}(e^ae^b)$ and
 \be
 \epsilon_{ij}^k=\begin{cases}
 -1\,,~~i,j<k\,,\\
 1\,,~~i,j\ge k\,,\\
 0~~\text{otherwise}\,.
 \end{cases}
 \ee

\section{From FNR to AKNS: transfer of the integrability structure}

\subsection{The map $\Psi_k$: From r-matrix structure of FNR to r-matrix structure of AKNS}\label{map_Psi}

Let $k\ge 1$ be fixed in all of this section. Our goal is to show that the $r$-matrix structure used to express the FNR equation \eqref{FNR_flows} is the origin of the $r$-matrix 
structure of the AKNS realization of the object $L^{(k)}=P_+(S^kL)$
. In view of Lemma \ref{lemma_shift}, this suggests that 
we should use the bracket $\{~,~\}_{-k}$ as a starting point. This is where our reformulation of the FNR flows in Proposition \ref{prop_flow2} becomes useful.

The following proposition is well-known in the literature (see e.g. \cite{FNR}). We formulate it in the form that motivates our subsequent analysis.
\begin{proposition}
\label{solving_FNR_eq}
Fix a time variable $t_k$. For $L\in\cL_-^0$,  
\be
L = \sum_{j=0}^\infty \ell_j\lda^{-j}\,,~~\ell_j=\mato 
a_j & b_j \\
c_j & -a_j
\matf\,,
\ee
Consider the flow equation
\be
\label{diff_FNR}
\frac{d}{dt_k}L=[L^{(k)},L]\,,
\ee
with periodic boundary conditions in $t_k$ and $\ell_0=\sigma_3$ \ie $a_0=1$, $b_0=0=c_0$. Then, one can view the elements $a_j$, $b_j$, $c_j$ as periodic functions of $t_k$ and
\begin{enumerate}
\item $b_j,c_j$, $j=1,\dots,k$ are not constrained.

\item $b_j,c_j$, $j>k$ are polynomials respectively denoted as $P_k^{b_j}$ and $P_k^{c_j}$ in $b_n^{(\ell)},c_n^{(\ell)}$, $n=1,\dots,k$ where $u_n^{(\ell)}$ is the $\ell$-th $t_k$ derivative of $u_n$, $u=b,c$. 

\item $a_j$, $j\ge 1$ are polynomials denoted as $Q_k^{a_j}$ in $b_n^{(\ell)},c_n^{(\ell)}$, $n=1,\dots,k$. For $j=0\dots,k$, $a_j$ is in fact a polynomial in $b_n^{(0)},c_n^{(0)}$, $n=0,\dots,k$ only, 
no derivatives appear.

\end{enumerate}
\end{proposition}
The last point on the form of $a_j$, $j=0\dots,k$ has a translation in Hamiltonian terms that we will exploit below. 
Denote now $V_k$ as the solution of the equation thus obtained.
We want to introduce a map $\Psi_k$ that conveniently captures the solution procedure of the previous proposition and which assigns the solution $V_k$ to $L$, 
\be
\Psi_k(L)=V_k\,.
\ee
$V_k$ is now a function of $t_k$ (and of course a power series in $\lda^{-1}$). We may sometimes write explicitly $V_k(\lda,t_k)$ to emphasize this point. 
The point is that after imposing the flow equation of Proposition \ref{solving_FNR_eq}, the variables $a_j$, $b_j$, $c_j$, which are free, $t_k$-independent coordinates on the coadjoint orbit, 
become $t_k$-dependent functions which are all expressible in terms of a finite number of functions $b_j(t_k),c_j(t_k)$, $j=1,\dots,k$ and their $t_k$ derivatives. The map $\Psi_k$ 
assigns to each coadjoint orbit variable precisely its expression after resolution of the constraints imposed by the flow equation \eqref{diff_FNR} with the chosen condition on $\ell_0$.
To formalise this, we need to define an algebra where $V_k$ lives.
\begin{definition}
\label{def_Bs}
\begin{enumerate}
\item[]

\item 
Let $\cB$ be the algebra over $\CC$ of polynomials in $a_i,b_i,c_i$, $i=0,1,2,\dots$ regarded as smooth complex-valued functions of the variable $t_k$. 

\item Let $\cB_n$ be the quotient of $\cB$ by the ideal $\cI_n$ generated by $\{a_i(t_k),b_i(t_k),c_i(t_k)\}_{i=n+1}^\infty$.  $\cB_n$ is identified with the 
associative algebra generated by $\{a_i(t_k),b_i(t_k),c_i(t_k)\}_{i=0}^n$.

\item Let $\cB^0_n$ be the algebra over $\CC$ of polynomials in $\{b_i(t_k),c_i(t_k)\}_{i=0}^{n}$.

\item Let $d\cB^0_n$ be the algebra over $\CC$ of polynomials in $\{b_i^{(\ell)}(t_k),c_i^{(\ell)}(t_k)\}_{(i,\ell)=(0,0)}^{(n,\infty)}$ where 
\bea
&&b_i^{(0)}(t_k)=b_i(t_k)\,,~~c_i^{(0)}(t_k)=c_i(t_k)\,,~~\\
&&\partial_k u_i^{(\ell)}(t_k)=u_i^{(\ell+1)}(t_k)\,,~~i=0,\dots,n\,,~~\ell\ge 0\,,~~u=b,c\,.
\eea
\end{enumerate}
\end{definition}

\begin{definition}
\label{map_Psi_k}
The map $\Psi_k$ from $\cL_-^0 $ to $\cL_{d\cB^0_k}$ is defined in two steps: first by giving its action on the elements $\{a_i,b_i,c_i\}_{i=0}^\infty$ of $L\in\cL_-^0$ as
\bea
\label{action1}
&&a_0\mapsto 1\,,~~b_0\mapsto 0\,,~~c_0\mapsto 0\\
&&(b_i,c_i)\mapsto (b_i(t_k),c_i(t_k))\,,~~i=1,\dots,k\\
\label{action2}
&&a_i\mapsto Q_k^{a_i}(b_{1}(t_k),c_{1}(t_k),\dots,b_{i-1}(t_k),c_{i-1}(t_k))\,,~~i=1,\dots,k\\
&&b_i\mapsto P_k^{b_i}(b^{(\ell)}_{1}(t_k),c^{(\ell)}_{1}(t_k),\dots,b^{(\ell)}_{i-1}(t_k),c^{(\ell)}_{i-1}(t_k))\,,~~i\ge k+1\\
&&c_i\mapsto P_k^{c_i}(b^{(\ell)}_{1}(t_k),c^{(\ell)}_{1}(t_k),\dots,b^{(\ell)}_{i-1}(t_k),c^{(\ell)}_{i-1}(t_k))\,,~~i\ge k+1\\
&&a_i\mapsto Q_k^{a_i}(b^{(\ell)}_{1}(t_k),c^{(\ell)}_{1}(t_k),\dots,b^{(\ell)}_{i-1}(t_k),c^{(\ell)}_{i-1}(t_k))\,,~~i\ge k+1
\eea
and second, by extending it componentwise to a morphism of algebras from $\cL_-^0$ to $\cL_{d\cB^0_k}$, the algebra of power series in $\lda^{-1}$ with coefficients in ${\rm sl}(2,d\cB^0_k)$. 
$P_k^{b_i}$, $P_k^{c_i}$ and $Q_k^{a_i}$ are the polynomials determined by the solution procedure of Proposition \ref{solving_FNR_eq}.
\end{definition}
\begin{remark} A geometric realization of the map $\Psi_k$ restricted to $\{a_i,b_i,c_i\}_{i=0}^k$ is given in Proposition 2 of \cite{R} 
(there $k=N$ and $t_N\equiv x$).
\end{remark}
\begin{remark} The motivation for the definition of the various algebras in Definition \ref{def_Bs} can be seen as follows. If we write
$V_k(\lda,t_k)=\Psi_k(L)$ as 
\be
V_k(\lda,t_k)=\sum_{j=0}^\infty V_{k,j}(t_k)\lda^{-j}\,,
\ee
then we see from Definition \ref{map_Psi_k} that for $j>k$, $V_{k,j}(t_k)$ belongs to ${\rm sl}(2,d\cB^0_k)$ and for $j=0,\dots,k$, it belongs to ${\rm sl}(2,\cB^0_k)$ (a subalgebra of ${\rm sl}(2,d\cB^0_k)$). We will only be interested in the Poisson properties of the generalized Lax matrix $V_k^{(k)}$ (see next Definition). The latter only involves $V_{k,j}(t_k)$, $j=0,\dots,k$ and hence, we will be concerned with a certain Poisson structure on $\cB_k^0$. The latter is most easily derived from a natural Poisson structure on $\cB$ and 
its subalgebra $\cB_k$.
\end{remark}

\begin{definition}
For n$\ge 0$, we define the (generalized) Lax matrices as
\be
V_k^{(n)}(\lda,t_k)=P_+(S^n(\Psi_k(L)))\,.
\ee
\end{definition}
For $k=1$, one recovers the well known Lax matrices of the ANKS hierarchy, with $b_1=q$ and $c_1=r$, the two fundamental fields in that hierarchy (more details 
in the Examples section below).

The map $\Psi_k$ is now used to transfer the $r$-matrix structure of the FNR theory over elements $L\in\cL_-^0$ to the Poisson structure of the Lax matrix 
$V_k^{(k)}$. This requires endowing $\cB$ and $\cB_n$ with natural Poisson structures such that the map $L\mapsto V_k^{(k)}(\lda,t_k)$
 enjoys nice Poisson properties. 
  To this end, note that the associative algebra $\cA$ over $\CC$  of polynomials in the symbols 
 $a_i,b_i,c_i$, $i=0,1,2,\dots$ inherits a Poisson structure from $\{~,~\}_{-k}$ on $\cL_-^0$ by viewing it as a subspace of $C^\infty(\cL_-^0)$ 
 (polynomials in those symbols are particular smooth functions in the entries of elements of $\cL_-^0$). In other words, it is enough to 
 specify the Poisson brackets of $a_i,b_i,c_i$ and the latter are obtained via the identifications
\be
\ell_i^+\mapsto b_i\,,~~\ell_i^-\mapsto c_i\,,~~\ell_i^3\mapsto a_i\,,
\ee
where we write an element $L$ of $\cL_-^0$ as 
\be
L=\sum_{j=0}^\infty \ell_j \lda^{-j}\,,~~\ell_j=\mato 
a_j & b_j \\
c_j & -a_j
\matf=\ell_j^+\sigma_++\ell_j^-\sigma_-+\ell_j^3\sigma_3\,,
\ee
and we have made a choice of basis in ${\rm sl}(2,\CC)$ corresponding to the three matrices 
\be
\label{basis}
\sigma_+=\mato 0& 1\\
0&0
\matf\,,~~\sigma_-=\mato 0&0 \\
1&0
\matf\,,~~\sigma_3=\mato 1& 0\\
0&-1
\matf\,.
\ee

Using \eqref{bracket_components}, we get:
\be
\label{PB_generators}
\{b_m,c_n\}_{-k}=-2\tilde{\epsilon}_{mn}^k a_{m+n-k-1}\,,~~
\{b_m,a_n\}_{-k}=\tilde{\epsilon}_{mn}^k b_{m+n-k-1}\,,~~
\{c_m,a_n\}_{-k}=-\tilde{\epsilon}_{mn}^k c_{m+n-k-1}\,,
\ee
where 
\be
 \tilde{\epsilon}_{mn}^k=\begin{cases}
 1\,,~~m,n>k\,,\\
 -1\,,~~m,n\le k\,,\\
 0~~\text{otherwise}\,.
 \end{cases}
 \ee

  The Poisson bracket on $\cA$ is extended componentwise to Laurent series with coefficients in ${\rm sl}(2,\cA)$.
$\cL_-^0$ is naturally embedded in the subalgebra $\cL^0_\cA$ of power series in $\lda^{-1}$ with coefficients in ${\rm sl}(2,\cA)$.

Similarly, $\cB$ can be viewed as the subalgebra in $C^\infty(\mathfrak{S})$ where $\mathfrak{S}$ is the space of smooth periodic maps from $S^1$  to 
$\cL_-^0$. Therefore, the passage from a Poisson structure on $\cA$ to one on $\cB$ uses a well-known procedure of central extension 
to incorporate the extra variable $t_k$ and the corresponding derivation $\partial_k$. The details of this construction, involving the 
so-called double loop algebra, can be found for instance in Lecture 7 of \cite{STS_lectures} for the case of the $R$-bracket. The generalization to the $k$-bracket was performed in \cite{RSTS} and 
this is the result we use here. We keep the same notation $\{~,~\}_{k}$ for the bracket resulting from the central extension construction. 

We first extend the bilinear form  to elements of $\mathfrak{S}$
as
\be
\label{ext_bilinear}
\left(X,Y\right)={\rm Res}_{\lda}\int{\rm Tr}(X(\lda,t_k)Y(\lda,t_k))\,dt_k\,.
\ee
One then introduces the following $2$-cocycle on $\mathfrak{S}$
\be
\omega(X,Y)=\int{\rm Tr}(X(\lda,t_k)\partial_k Y(\lda,t_k))\,dt_k
\ee
and the corresponding central extension defined as the Lie algebra of pairs $(X,p)$ where $X\in\mathfrak{S}$ and $p\in \CC(\lda,\lda^{-1})$ endowed with the Lie 
bracket
\be
\label{extended_bracket}
[(X,p),(Y,q)]=([X,Y],\omega(X,Y))\,.
\ee
The natural extension of the bilinear form on $\mathfrak{S}$ to its central extension is 
\be
\langle (X,p),(Y,q)\rangle=(X,Y)+{\rm Res}_{\lda}p(\lda)q(\lda)\,.
\ee
We now use a well-known result:
\begin{proposition}
If $R$ satisfies the modified classical Yang-Baxter equation and $\omega$ is a $2$-cocycle with respect to the Lie bracket $[~,~]$ then $\omega_R$ 
defined by
\be
\omega_R(X,Y)=\frac{1}{2}\omega(RX,Y)+\frac{1}{2}\omega(X,RY)
\ee
is a $2$-cocycle with respect to the Lie bracket $[~,~]_R$.
\end{proposition}
This allows one to define a central extension of $\mathfrak{S}$ with respect to $\omega_R$ as the Lie algebra of pairs $(X,p)$ equipped with the following bracket
\be
[(X,p),(Y,q)]_R=([X,Y]_R,\omega_R(X,Y))\,.
\ee
It can be viewed consistently as the $R$-bracket associated to $[~,~]$ in \eqref{extended_bracket} if we define the action of $R$ on $(X,p)$ as
\be
R(X,p)=(RX,p)\,.
\ee
This construction applies to any solution of the classical Yang-Baxter equation. In particular, it applies to $R\circ S^k$. We need to specify the action of the intertwining operator $S^k$ on $(X,p)$ as
\be
S^k(X,p)=(S^kX,s^kp)\,,~~(s^kp)(\lda)=\lda^kp(\lda)\,.
\ee
A simple but important observation is that the Lie bracket \eqref{extended_bracket} does not depend on $p$ or $q$. This is what allows one 
to define a Poisson bracket on smooth functions on $\mathfrak{S}$ starting from the standard construction of a Lie-Poisson bracket for functions 
on the central extension of $\mathfrak{S}$. More precisely, we define, for $k\in\ZZ$,
\be
\{F,G\}_k^p(L)=\langle (L,p),[(\nabla F(L),0),(\nabla G(L),0)]_{R\circ S^k}\rangle\,,~~\forall L\in \mathfrak{S}\,,~~\forall F,G\in C^\infty(\mathfrak{S})\,,
\ee
where the gradient of $F\in C^\infty(\mathfrak{S})$ is calculated with respect to \eqref{ext_bilinear} now. The following proposition gives 
explicit formulas for this bracket that are useful in our subsequent analysis.
\begin{proposition}
We have
\be
\label{extended_PB}
\{F,G\}_n^p(L)=-\frac{1}{2}\left( [RS^n\nabla F(L),L]-S^nR[\nabla F(L),L]+(pRS^n-S^nRp)\partial_k\nabla F(L)),\nabla G(L) \right)
\ee
where one uses the bilinear form \eqref{ext_bilinear} now,
and the generalization of 
\eqref{bracket_components} to $\mathfrak{S}$ reads \cite{RSTS}, with our 
conventions, 
\be
 \{L_i^a(t_k),L_j^b(\tau_k)\}^p_k=\epsilon_{ij}^k(K^{-1})_{ac}\,C^{cd}_b\,L^d_{i+j+1-k}(t_k)\delta(t_k-\tau_k)-\epsilon_{ij}^k(K^{-1})_{ab}
 \delta_{i+j+1-k,s}p_s\delta^{'}(t_k-\tau_k)\,,
 \ee
 where $p=p_s\lda^s$ and we have denoted 
 \be
  \{L_i^a(t_k),L_j^b(\tau_k)\}^p_k\equiv  \{f_i^a(t_k),f_j^b(\tau_k)\}^p_k
 \ee
 where $f_i^a(t_k)$ is the coordinate function $f_i^a(t_k):L\mapsto L_i^a(t_k)$ if $\displaystyle L(\lda,t_k)=\sum_{i=-\infty}^0L_i^a(t_k)e^a\lda^i$.
 \end{proposition}
 \prf
 The first equation is obtained by noting that 
 \be
 \langle (Y,p),[(X,0),(Z,0)]\rangle=-([X,Y]+p\partial_k X,Z)\,,~~\forall X,Y,Z\in \mathfrak{S}\,,
 \ee
 and 
 \be
 (RS^nX,Y)=-(X,S^nRY)\,,~~\forall n\in\ZZ\,,~~\forall X,Y\in \mathfrak{S}\,.
 \ee
 The coordinate expression is obtained by direct calculation from the first equation using the equality
 \be
 \nabla f_i^a(t_k)(L)=(K^{-1})_{ac}e^c\lda^{-i-1}\delta_{t_k}\,,~~\delta_{t_k}(\theta)=\delta(t_k-\theta)~~\text{(Dirac distribution)}\,.
 \ee

\noindent {\bf Remark:} The appearance of  $K^{-1}$ compared to \cite{RSTS} is due to our implicit choice of identifying the loop algebra and its dual from the beginning using the ad-invariant bilinear form $(~,~)$.

The bracket $\{~,~\}_k^p$ enjoys the following property which generalizes Lemma \ref{lemma_shift}.
\begin{lemma}
\label{lemma_shift_ext}
 For all $k\in\ZZ$ and $p\in\CC(\lda)$,
 \be
\{F,G\}_k^{s^kp}(S^kL)=\{F\circ S^k,G\circ S^k\}_R^p(L)\,~~\forall L\in\mathfrak{S}\,~~\forall F,G\in C^\infty(\mathfrak{S})\,~~
\ee
\end{lemma}

\noindent {\bf Notation:} In practice, we will use only functions $p$ of the form $p(\lda)=\lda^r$ for a given $r\in \ZZ$ so we will denote the corresponding $k$-bracket 
by $\{~,~\}^r_k$. We have a family of Poisson brackets that are labeled by two integers $k$ and $r$. To simplify even further the notations, 
we will simply write $\{~,~\}_k$ for $\{~,~\}^k_k$, \ie when $r=k$.

\begin{proposition}\label{subalgebra}
$\cB_k$ is a Poisson subalgebra of $\cB$ equipped with the Poisson bracket $\{~,~\}_{-k}$.
\end{proposition}
\prf
We need to check that the ideal $\cI_k$ is also a Poisson ideal for $\{~,~\}_{-k}$. 
Recall that
\be
\mato
a_j(t_k) & b_j(t_k) \\
c_j(t_k) & -a_j(t_k)
\matf=\ell_j(t_k)=L_{-j}(t_k)\,,~~j\ge 0\,,
\ee
so 
\be
 \{\ell_i^a(t_k),\ell_j^b(\tau_k)\}_{-k}=\epsilon_{-i,-j}^{-k}(K^{-1})_{ac}\,C^{cd}_b\,\ell^d_{i+j-1-k}(t_k)\delta(t_k-\tau_k)-\epsilon_{-i,-j}^{-k}(K^{-1})_{ab}
 \delta_{i+j-1-k,k}\delta^{'}(t_k-\tau_k)\,.
 \label{PBS}
\ee
Hence, for $0\le i\le k$ and $j\ge k+1$ we get
\be
 \{\ell_i^a(t_k),\ell_j^b(\tau_k)\}_{-k}=0\,\in \cI_k\,,
\ee
and for $i,j\ge k+1$,
\be
 \{\ell_i^a(t_k),\ell_j^b(\tau_k)\}_{-k}=-(K^{-1})_{ac}\,C^{cd}_b\,\ell^d_{i+j-1-k}(t_k)\delta(t_k-\tau_k) \,   \in\cI_k
\ee
where the last claim follows from $i,j\ge k+1\Rightarrow i+j-k-1\ge k+1$.
\finprf

In practice, this means that we can restrict $\{~,~\}_{-k}$ to the Poisson subspace of elements of the form
\be
\Lambda^{(k)}(t_k)=\sum_{j=0}^k\mato 
a_j(t_k) & b_j(t_k) \\
c_j(t_k) & -a_j(t_k)
\matf \lda^{-j}\equiv \sum_{j=0}^k \ell_j(t_k)\lda^{-j}=S^{-k}P_+(S^k(L(t_k)))
\ee

$\Lambda^{(k)}(t_k)$ is thus simply a truncation of $L(t_k)$ to its first $k+1$ terms.

As can be seen from \eqref{action1}-\eqref{action2}, for any $n$ such that $0\le n\le k$, the map $\Psi_k$ restricted to $\cB_n$, acts trivially on $\{b_i,c_i\}_{i=0}^n$ and substitutes for $a_i$ certain polynomials in $b_j,c_j$, $j< i$.
On each element $\displaystyle \Lambda^{(n)}(t_k)=
\sum_{j=0}^n \ell_j(t_k) \lda^{-j}$ we can define the analog of the Hamiltonian functions \eqref{Casimirs} of the FNR theory which we also denote by $\phi_j$,
\be
\phi_j(\Lambda^{(n)}(t_k))=-\frac{1}{2}{\rm Res}_{\lda}\,\lda^j{\rm Tr}(\Lambda^{(n)}(t_k))^2\,.
\ee
It turns out that the action of $\Psi_k$ on $a_i$ coincides precisely to fixing $\phi_j(\Lambda^{(n)}(t_k))$, $j=0,\dots,n-1$ to constant values. This is a well-known consequence of solving 
\eqref{diff_FNR}. It is most often stated as the fact that in the equation for $a_j$ in \eqref{diff_FNR}
\be
\partial_k a_j= P(a_i,b_i,c_i)
\ee
the right-hand side is always a total $t_k$ derivative. Therefore one can always integrate the equation up to a constant of integration. In the particular case where $j=1,\dots,n$, $n\le k$, 
the constant of integration is half the constant value of $\phi_{j-1}$.
One of the crucial aspects of our construction is that, from the point of view of the Poisson structure $\{~,~\}_{-k}$, this procedure of reducing the phase space of the theory by fixing the value of the functions $\phi_i$ is consistent and the Poisson structure then projects down to the reduced phase space. We have
\begin{proposition}\label{prop_Casimirs}
The functions $\phi_j$, $j=0,\dots,n-1$ are Casimir functions on $\cB_n$ equipped with $\{~,~\}_{-n}$ for any $n\le k$.
\end{proposition}

It is interesting to give two proofs of this fundamental statement using respectively the $r$-matrix and $R$ operator formalism.

\prf ($r$-matrix)

This proof runs along the following steps: 

Consider first the Poisson bracket structure $\{~,~\}_{-n} $without $\delta'$ terms.

{\bf Step 1:} The functions $\phi_j$ indexed from $0$ to $n-1$ are identified with the $n$ first powers of the Trace of $L^2(t_k)$ (from 0 to $-n+1$) since the truncation of $L$ 
to $\Lambda$ does not modify the $n$ first terms of the squared trace. They are thus trivially also the $n$ first powers (from $n+1$ to $2n$) of the Trace of $(\lambda^n L)^2(t_k)$ 

{\bf Step 2:} The Poisson bracket structure $\{ ~,~\}_{-n} $ evaluated on $(\lambda^n L)(t_k) \equiv L^n $ (not to be confused with its projected image $L^{(n)}$) 
is again parametrized by the skew-symmetric canonical $r$-matrix. We now incorporate the distribution $\delta(x-y)$ into the definition of $r$ :
\be
 \{L_1^n(\lda,x),L_2^n(\mu,y)\}_{-n}=-\frac{1}{2}[r(\lda,\mu),L^n(\lda,x)\otimes \1]+\frac{1}{2}[r(\mu,\lda),\1\otimes L^n(\mu,y)]\,,
 \ee
 where again $\Pi$ is the permutation operator on two copies of $\cL$: $\Pi(X\otimes Y)=Y\otimes X$ and 
\be
r(\lda,\mu)=-2\frac{1}{\lda-\mu}t \,\delta(x-y)\,,~~t=\Pi^{sl(2,\CC)}
\ee
This now identifies $\{~,~\}_{-n}$ acting on $(\lambda^n L)(t_k) \equiv L^n \equiv S^n L(t_k)$ with the original Poisson bracket $\{~,~ \}_R $.

{\bf Step 3:} Compute now the Poisson bracket of Trace of $(L^n)^2(\lda, t_k)$ with $L^n(\mu)$. By cyclicity of trace and properties of the $\Pi$ operator it yields
$\displaystyle \frac{[L^n(\lambda), L^n(\mu)]}{\lda - \mu} $. Counting now available powers of $\lambda$ indicates that no power higher than $n$ occurs in this expression whereas the Poisson bracket of the $n$ higher $\lambda$ powers of $(\lambda^n L)^2(t_k)$ with $L^n(\mu)$ should yield powers of $\lda$ running from $n$ to $2n$. This Poisson bracket is therefore null and the statement is proven. The functions $\phi_j$ are actually even Casimirs on $\cB$. 

Consider now the contribution of the $\delta'$ terms. They do not contribute to the Poisson brackets of the $n$ first terms of the matrix $L$ due to the term $\delta_{i+j-1-n,n}$ in the centrally extended
Poisson structure \eqref{PBS}. The statement that the functions $\phi_j$, $j=0,\dots,n-1$ be Casimir functions thus holds at least when Poisson-bracketed with elements of $\cB_n$. It actually holds indeed for $\cB$ as seen now from the second proof:
\finprf
\prf ($R$-operator)
Given that
\be
\nabla \phi_j(L(t_k))=-S^j (L)\delta_{t_k}\,,
\ee
a direct calculation from \eqref{extended_PB} yields, 
\bea
\{\phi_j,G\}_{-n}(L(t_k))&=&-\frac{1}{2}\left( ([RS^{j-n}L,L]+S^{-n}R[S^jL,L])\delta_{t_k},\nabla G(L) \right)\\
&&+\frac{1}{2}\left( (pRS^{-n}-S^{-n}Rp)S^jL \delta^{'}_{t_k},\nabla G(L) \right)\,.
\eea
Remembering that $j<n$, we have $RS^{j-n}L=-S^{j-n}L$. Hence the first line is zero since $[S^mL,L]=0$ for all $m\in\ZZ$. 
The second line is also seen to be zero by noting that, since $p(\lda)=\lda^{-n}$  (see the remark on notations after Lemma \ref{lemma_shift_ext}), we have for all $X\in\mathfrak{S}$,
\be
(pRS^{-n}-S^{-n}Rp)X=(S^{-n}RS^{-n}-S^{-n}RS^{-n})X=0\,,~~p(\lda)=\lda^{-n}\,.
\ee
We have actually shown more than announced in the Proposition since we see that $\phi_j$, $j=0,\dots,n-1$ is in fact a Casimir on the whole algebra $\cB$ 
equipped with $\{~,~\}_{-n}$. By restriction, it is also true on the subalgebra $\cB_n$.
\finprf
The functions $\phi_j$ being Casimirs, we can fix them to a definite value and obtain reduced Poisson manifolds as the level sets of $\phi_j$, $,=1,\dots,n-1$.
This proposition is particularly useful when $n=k$ since in that case we know that $\cB_k$ is a Poisson subalgebra of $\cB$. Combined with 
the previous observation which allows us to further restrict the Poisson bracket to the submanifold parametrized by $\phi_j(\Lambda^{(n)}(t_k))=cst$, 
$j=0,\dots,k-1$, this means that the elements of the form 
\be
S^{-k}P_+(S^k(\Psi_k(L)))=\sum_{j=0}^k\mato 
Q_k^{a_j} & b_j(t_k) \\
c_j(t_k) & -Q_k^{a_j}
\matf \lda^{-j}
\ee
form a Poisson submanifold of $\cL_{\cB_k^0}$ equipped with $\{~,~\}_{-k}$, where $Q_k^{a_j}$ is the polynomial substituted to $a_j$ by $\Psi_k$ (cf Definition \ref{map_Psi_k}).
We are now ready to state the main theorem of this section. We use the auxiliary space notation \eqref{aux_space} e.g.
\be
V_{k,1}^{(k)}(\lda,t_k)=V_{k}^{(k)}(\lda,t_k)\otimes \1\,.
\ee
\begin{theorem}
\label{theorem_Poisson_algebra}
The map $P_+\circ S^k\circ\Psi_k:L\mapsto V_k^{(k)}(\lda,t_k)$ is a Poisson map from $(\cL_-^0,\{~,~\}_{-k})$ to $(\cL_{\cB_k^0},\{~,~\}_R)$.
As a consequence, the Lax matrix $V_k^{(k)}(\lda,t_k)$ satisfies the following ultralocal Poisson algebra
\be
\label{Lax_matrix_bracket}
\{V_{k,1}^{(k)}(\lda,t_k),V_{k,2}^{(k)}(\mu,\tau_k)\}_{R}=\delta(t_k-\tau_k)[r(\lda,\mu),V_{k,1}^{(k)}(\lda,t_k)+V_{k,2}^{(k)}(\mu,\tau_k)]
\ee
where $r(\lda,\mu)$ is the characteristic ${\rm sl}(2,\CC)$ classical $r$-matrix of the AKNS hierarchy. The underlying labels $1,2$ in $V_k$ take the same meaning of ``position in the tensor
square of $\cL$'', as described in e.g. (\ref{TPB}).
\end{theorem}
\prf

To prove the first part of the statement, note that the map $\Phi_k\equiv S^{-k}\circ P_+\circ S^k\circ\Psi_k$ is a Poisson map
with respect to $\{~,~\}_{-k}$. This is a consequence of Propositions \ref{subalgebra}, \ref{prop_Casimirs} and the discussion following them. Indeed, 
Proposition \ref{subalgebra} states that the quotient map $\cB\to\cB/\cI_k\cong\cB_k$ is a Poisson map. This implies that the map $\cL_-^0\to \cL_{\cA_k}$
which acts as $L\mapsto S^{-k}P_+S^k(L)\equiv \Lambda^{(k)}$ is also a Poisson map\footnote{Note that in this intermediate step, we work with 
the algebra $\cA_k$ defined from $\cA$ in the same way as $\cB_k$ is defined from $\cB$. This is because the map $\Psi_k$ has not been applied yet so the elements $a_i$, $b_i$, $c_i$ 
are not yet viewed as functions of $t_k$. In practice, this simply means that we work with the Poisson structure based on the $r$-matrix before central extension.
This has no bearings on the arguments which only depend on the algebraic structure of the $r$-matrix and not on the details of the central extension, 
as the reader can check.}. Now,
\be
S^{-k}P_+S^k\Psi_k(L)=\Psi_k S^{-k}P_+S^k(L)\,.
\ee
In the right-hand side, $\Psi_k$ acts as the identity on $b_j,c_j$, $j=1,\dots,k$ and exactly as the fixing of the Casimirs $\phi_j$, $j=0,\dots,k-1$ 
to constants (say zero), on
$a_p$, $p=1,\dots,k$ with the effect of replacing $a_p$ by the polynomial $Q_k^{a_p}$. Proposition \ref{prop_Casimirs} thus ensures that it is a 
Poisson map from $\cL_{\cA_k}$ to $\cL_{\cB_k^0}$\footnote{See previous footnote.}. Putting everything together 
yields the claim on $\Phi_k$.
To conclude, it suffices to note that $P_+\circ S^k\circ\Psi_k=S^{k}\circ \Phi_k$ and that $S^k$ maps the bracket $\{~,~\}_{-k}$ to 
the $R$-bracket according to Lemma \ref{lemma_shift_ext}. 

The second part of the statement is a consequence of the Poisson map property combined with the classical $r$-matrix representation of the $k$-bracket 
discussed in Proposition \ref{Sklyanin_formula}, properly extended to functions of $t_k$ according to the central extension construction 
explained above (see also \cite{RSTS}). 
For the $R$-bracket, the latter reads generically
\be
\label{sklyanin_delta}
\{L_1(\lda,t_k),L_2(\mu,\tau_k)\}_R=\delta(t_k-\tau_k)[\frac{\Pi}{\lda-\mu},L_1(\lda)+L_2(\mu)]\,.
 \ee
The equality 
\be
\{F\circ S^k\circ\Phi_k,G\circ S^k\circ\Phi_k\}_{-k}(L(t_k))=\{F,G\}_{R}(S^k\circ\Phi_k(L(t_k)))
\ee 
ensures that we can consistently restrict \eqref{sklyanin_delta} to $S^k\circ\Phi_k(L(t_k)=V_k^{(k)}(\lda,t_k)$ to obtain \eqref{Lax_matrix_bracket} as claimed.
\finprf

\noindent {\bf Remark:} This theorem proves in full generality the conjecture made in our previous paper \cite{ACDK} where we had explicitly established 
\eqref{Lax_matrix_bracket} for $k=1,2,3$. We had used a completely different approach based on a Lagrangian and ideas of covariant field theory and presented 
arguments in favour of the claim that \eqref{Lax_matrix_bracket} holds beyond $k=3$.

\subsection{From FNR flows to AKNS Lax operators}\label{FNRtoAKNS}

To establish consistency of the $\Psi_k$ constraint procedure with integrability structures we now consider the construction of the time-flow operators in the AKNS hierarchy, as derived from traces of the monodromy of the Lax `space'' operator, and compare it with the application of the constraint onto the time-flow matrices in the FNR procedure.
We will see that both objects are identical. More precisely, the generalized Lax matrices $V_k^{(n)}(\lda,t_k)$, $n\in\NN$ obtained by applying $\Psi_k$ in the FNR picture should coincide with 
the matrices $\cV_k^{(n)}(\lda,t_k)$ constructed from the monodromy of $V_k^{(k)}(\lda,t_k)$ at the AKNS level and which  
generate the $t_n$ flow of $V_k^{(k)}(\lda,t_k)$ with respect to $\{~,~\}_R$.
The proof runs in three steps.

{\bf First step:} 

The commutativity of the flows at the FNR level implies the (weak) zero curvature condition
\be
\label{weak_ZC}
[\partial_n L^{(k)}-\partial_k L^{(n)}+[L^{(k)},L^{(n)}],L]=0
\ee
for any pair of times $t_k,t_n$, where we recall that $L^{(n)}=P_+(S^nL)$. In fact, one has the stronger result
\begin{lemma}
The FNR equations
\be
L_{t_k}=[L^{(k)},L]=\frac{1}{2}[RS^kL,L]\,,~~k\ge 1\,,
\ee
imply the (strong) zero curvature condition:
\be
\label{strong_ZC}
\partial_n L^{(k)}-\partial_k L^{(n)}+[L^{(k)},L^{(n)}]=0\,.
\ee
\end{lemma}
We propose again two distinct proofs of this lemma.

\prf
The proof relies on the following facts:
\begin{enumerate}
\item $R=P_+-P_-$ satisfies the modified classical Yang-Baxter equation
\be
[RX,RY]-2R[X,Y]_R=-[X,Y]\,,
\ee

\item $S^j$ is an intertwining operator,

\item $R$ and $S^j$ commute with all the differential operators $\partial_{t_k}$, $k\ge 1$.
\end{enumerate}
Now we have on the one hand
\bea
(RS^nL)_{t_k}-(RS^kL)_{t_n}+\frac{1}{2}[RS^nL,RS^kL]&=&\frac{1}{2}RS^n[RS^kL,L]-\frac{1}{2}RS^k[RS^nL,L]+\frac{1}{2}[RS^nL,RS^kL]\nonumber\\
&=&R[S^kL,S^nL]_R-\frac{1}{2}[RS^kL,RS^nL]\nonumber\\
&=&\frac{1}{2}[S^kL,S^nL]\nonumber\\
&=&0\nonumber
\eea
and on the other hand, using $R=2P_+-I$,
\bea
(RS^nL)_{t_k}-(RS^kL)_{t_n}+\frac{1}{2}[RS^nL,RS^kL]&=&2\left(L^{(n)}_{t_k}-L^{(k)}_{t_n}+[L^{(n)}_{t_k},L^{(k)}_{t_n}]\right)\nonumber\\
&&-S^n\left(L_{t_k}-[L^{(k)},L]\right)\nonumber\\
&&+S^k\left(L_{t_n}-[L^{(n)},L] \right)\nonumber\\
&=&2\left(L^{(n)}_{t_k}-L^{(k)}_{t_n}+[L^{(n)}_{t_k},L^{(k)}_{t_n}]\right)\,.
\eea
\finprf
A more compact but less rigorous proof is given as follows:

\prf
Equation \eqref{weak_ZC} implies that $\partial_n L^{(k)}-\partial_k L^{(n)}+[L^{(k)},L^{(n)}]$ is a matrix commuting with $L$. Hence it must reduce in general to a polynomial in $L$ hereafter 
denoted $P(L)$. However
$\partial_n L^{(k)}-\partial_k L^{(n)}+[L^{(k)},L^{(n)}] $ is at most of order $n+k$ as a polynomial in $\lambda$ whilst any polynomial in $L$ must exhibit 
all finite powers of $\lda$. Moreover $P(L)$ cannot be reduced to the identity matrix since $\partial_n L^{(k)}-\partial_k L^{(n)}+[L^{(k)},L^{(n)}]$
is traceless. This leaves us with $P(L)=0$, which is \eqref{strong_ZC}.
\finprf

{\bf Second step:}

We fix a time $t_k$ and assume that one has solved the FNR equation with respect to $t_k$ first, producing the map $\Psi_k$. The application 
of $\Psi_k$ to \eqref{strong_ZC} yields the AKNS zero-curvature condition
\be
\partial_n V_k^{(k)}(\lda,t_k)-\partial_k V_k^{(n)}(\lda,t_k)+[V_k^{(k)}(\lda,t_k),V_k^{(n)}(\lda,t_k)]=0\,,
\ee
for any other time $t_n$, where Lax matrices should now be viewed as functions of $t_n$ as well. 
The collection of 
these equations can be cast into a single equation using the following generating function 
\be
V_k(\lda,\mu,t_k)=\sum_{n=1}^\infty \mu^{-n}V_k^{(n-1)}(\lda,t_k)=\sum_{n=1}^\infty \mu^{-n}P_+(S^{n-1}(\Psi_k(L(t_k))))=
\sum_{n=1}^\infty \mu^{-n}P_+(\lda^{n-1} V_k(\lda,t_k))
\ee
and the differential operator $\displaystyle D=\sum_{n=1}^\infty \mu^{-n}\partial_{n-1}$,
\be
D V_k^{(k)}(\lda,t_k)-\partial_k V_k(\lda,\mu,t_k)+[V_k^{(k)}(\lda,t_k),V_k(\lda,\mu,t_k)]=0\,.
\ee

On the other hand, at the AKNS level, the following generalization of a classic argument (see e.g. \cite{FT} for an $r$-matrix derivation or 
\cite{KR} for a geometric treatment) was proved in \cite{ACDK}. Denote $\xi=t_k$ and assume 
$\xi\in[-M,M]$ (with periodic boundary conditions). Let $T(\xi_1,\xi_2,\lda)$ be the finite-interval monodromy of the auxiliary problem
\be
\label{pb_aux}
\partial_\xi \Psi(\lda,\xi)=V_k^{(k)}(\lda,\xi)\Psi(\lda,\xi)
\ee
normalised to $T(\xi,\xi,\lda)=\1$. Introduce the following decomposition, understood as a series in $\lda^{-1}$,
\be
\label{decomposition}
T(\xi_1,\xi_2,\lda)=(\1+W(\xi_1,\lda))e^{Z(\xi_1,\xi_2,\lda)}(\1+W(\xi_2,\lda))^{-1}
\ee
where $W$ is off-diagonal and $Z$ is proportional to $\sigma_3$ and admits the expansion
\be
Z(\xi_1,\xi_2,\lda)=\sum_{n=0}^k Z^{(-n)}(\xi_1,\xi_2)\lda^n+\sum_{n=1}^\infty Z^{(n)}(\xi_1,\xi_2)\lda^{-n}\,.
\ee
Define 
\be
\label{def_Ham}
H_k^{(n-1)}=\frac{1}{2}{\rm Tr}(\sigma_3Z^{(n)}(-M,M))\,,~~n\ge 1\,.
\ee
The functions $H_k^{(n)}$ are thus commuting Hamiltonians which respectively generate the $t_n$ flows with respect to $\{~,~\}_R$. We have 
\begin{proposition}\cite{ACDK}
\label{Hamiltonian_ZC}
The Hamiltonian flow equation
\be
\partial_n V_k^{(k)}(\lda,t_k)=\{V_k^{(k)}(\lda,t_k),H_k^{(n)}\}_R
\ee
is equivalent to the zero curvature equation
\be
\partial_n V_k^{(k)}(\lda,t_k)-\partial_k\cV_k^{(n)}(\lda,t_k)+[V_k^{(k)}(\lda,t_k),\cV_k^{(n)}(\lda,t_k)]=0
\ee
where $\cV_k^{(n)}(\lda,t_k)$ is obtained from the generating function
\be
\cV_k(\lda,\mu,\xi)=\frac{-1}{\lda-\mu}(\1+W(\xi,\mu))\sigma_3(\1+W(\xi,\mu))^{-1}
\ee
via the expansion
\be
\cV_k(\lda,\mu,\xi)=\sum_{n=0}^\infty \frac{\cV_k^{(n)}(\lda,\xi)}{\mu^{n+1}}\,.
\ee
\end{proposition}

{\bf Third step:}

We now have two candidate generating functions for what is expected to be the same hierarchy of Lax matrices. This is consistent if they are equal.
Indeed we prove:

\begin{theorem}
\label{equality_V}
The two generating functions $V_k(\lda,\mu,t_k)$ and $\cV_k(\lda,\mu,t_k)$, obtained either directly from the FNR level by application of the map $\Psi_k$; or 
in a self-contained manner at the AKNS level as explained above; are equal, and therefore:
\be
V_k^{(n)}(\lda,t_k)=\cV_k^{(n)}(\lda,t_k)\,,~~n\ge 0\,.
\ee
\end{theorem}
\prf
On the one hand, 
\bea
V_k(\lda,\mu,t_k)&=&\sum_{n=1}^\infty \mu^{-n}P_+(\lda^{n-1} V_k(\lda,t_k))\\
&=&\sum_{n=0}^\infty \mu^{-n-1}P_+(\lda^n\sum_{j=0}^\infty V_{k,j}(t_k)\lda^{-j})\\
&=&\frac{1}{\mu}\sum_{n=0}^\infty\sum_{\ell=0}^n \mu^{\ell-n} \frac{\lda^{\ell}}{\mu^\ell} V_{k,n-\ell}(t_k) \\
&=&\frac{1}{\mu}\sum_{\ell=0}^\infty\frac{\lda^{\ell}}{\mu^\ell}\sum_{n=\ell}^\infty\mu^{\ell-n} V_{k,n-\ell}(t_k)\\
&=&\frac{-1}{\lda-\mu}\sum_{p=0}^\infty \mu^{-p} V_{k,p}(t_k)\\
&=&\frac{-1}{\lda-\mu}V_k(\mu,t_k)
\eea
and by definition of $V_k(\mu,t_k)$, we have
\be
\partial_k V_k(\mu,t_k)=[V_k^{(k)}(\mu,t_k),V_k(\mu,t_k)]\,.
\ee
On the other hand
\be
\cV_k(\lda,\mu,t_k)=\frac{-1}{\lda-\mu}\cV_k(\mu,t_k)\,.
\ee
where 
\be
\cV_k(\mu,t_k)=(\1+W(t_k,\mu))\sigma_3(\1+W(t_k,\mu))^{-1}\,.
\ee
Hence,
\bea
\label{temp1}
\partial_k\cV_k(\mu,t_k)=[\partial_k W(\mu,t_k)(\1+W(\mu,t_k))^{-1},\cV_k(\mu,t_k)]
\eea
Now, we use the fact that $T(\xi_1,\xi_2,\lda)$ in \eqref{decomposition} is a solution of \eqref{pb_aux} for $\xi_1=\xi=t_k$ and $\xi_2$ 
arbitrary. We obtain, dropping $(\mu,t_k)$ for conciseness,
\be
\partial_k W(\1+W)^{-1}=V_k^{(k)}-(\1+W)\partial_k Z(\1+W)^{-1}\,.
\ee 
Inserting in \eqref{temp1} and remembering that $Z$ is proportional to $\sigma_3$, so that the contribution 
$(\1+W)\left[\partial_k Z,\sigma_3\right](\1+W)^{-1}$ is zero, we are left with 
\bea
\partial_k\cV_k(\mu,t_k)=[V_k^{(k)}(\mu,t_k),\cV_k(\mu,t_k)]
\eea
which is the same equation as for $V_k(\mu,t_k)$. 
Solutions $\Upsilon$ of this equation, known as resolvents, have been studied extensively (see e.g. \cite{Dickey_book}, Chap. 9 and 10). 
They are known to form a two-dimensional vector space spanned by $\Upsilon^\alpha=\phi\,E_{\alpha\alpha}\,\phi^{-1}$, $\alpha=1,2$ where $\phi$ is a so-called 
dressing factor realising a gauge transformation from $V_k^{(k)}$ to a diagonal matrix $D(\mu)$ and $E_{\alpha\alpha}$ is the matrix with the only non zero entry equal to $1$ 
at position $(\alpha,\alpha)$. The dressing $\phi$ can be chosen in the form $\1+W$ as we have here.
Then a resolvent is completely characterised by fixing its constant coefficient (in front of $\mu^0$). Here both $\cV_k(\mu,t_k)$ and $V_k(\mu,t_k)$ 
have constant coefficient equal to $\sigma_3$. Therefore, they are both equal to 
\be
(\1+W(t_k,\mu))\,\sigma_3\,(\1+W(t_k,\mu))^{-1}
\ee
as required.
\finprf

\noindent {\bf Remark:} To obtain this equality, we have chosen to set all the constants of integration in the map $\Psi_k$ to zero so that diagonal elements $a_j$, $j\ge 1$ 
contain no constant term. This is the standard ANKS normalisation.

\subsection{Zero curvature representation and dual Hamiltonian formulation of an integrable PDE}\label{dualHam}

To complete our diagram on Figure 1, we must show that if we pick two times $t_k$ and $t_n$, applying first $\Psi_k$ and obtaining
the zero curvature associated to $t_n$ or, applying first $\Psi_n$ and 
obtaining the zero curvature associated to $t_k$, yields the same set of PDEs for the same variables. Without loss of generality, fix 
$1\le n < k$. We begin with a simple but convenient observation.

\begin{lemma}
Considering the matrices $\ell_j$ as functions of both $t_n$ and $t_k$, applying $\Psi_n$ to obtain the 
generalized Lax matrices $V_n^{(j)}$ and imposing the zero curvature condition
\be
\label{ZCC_V}
\partial_kV_n^{(n)}-\partial_nV_n^{(k)}+[V_n^{(n)},V_n^{(k)}]=0
\ee
is equivalent to imposing the following set of simultaneous equations 
\bea
\label{FNR_comp}
&&\ell_0=\sigma_3\,,~~\partial_n\ell_p=\sum_{j=0}^n[\ell_j,\ell_{p+n-j}]\,,~~p\ge 1\,,\\
\label{ZCC_L}
&&\partial_kL^{(n)}-\partial_nL^{(k)}+[L^{(n)},L^{(k)}]=0\,,~~L^{(j)}=\sum_{m=0}^j\ell_m\lda^{j-m}\,.
\eea
\end{lemma}
\prf
Eq \eqref{FNR_comp} is just the component form of the FNR equation \eqref{diff_FNR} of Proposition \ref{solving_FNR_eq} which gives rise to the map $\Psi_n$. 
Hence, under \eqref{FNR_comp}, the zero curvature condition \eqref{ZCC_L} is equivalent to \eqref{ZCC_V}.
\finprf
We are now in a position to prove
\begin{proposition}\label{equivalence_nk}
Let $1\le n <k$. The set of equations
\bea
\label{FNR_comp_n}
&&\ell_0=\sigma_3\,,~~\partial_n\ell_p=\sum_{j=0}^n[\ell_j,\ell_{p+n-j}]\,,~~p\ge 1\,,\\
\label{ZCC_L_n}
&&\partial_kL^{(n)}-\partial_nL^{(k)}+[L^{(n)},L^{(k)}]=0\,.
\eea
gives rise to the same set of PDEs for the variable $b_j,c_j$, j$=1,\dots,n$ as the set of equations
\bea
\label{FNR_comp_k}
&&\ell_0=\sigma_3\,,~~\partial_k\ell_p=\sum_{j=0}^k[\ell_j,\ell_{p+k-j}]\,,~~p\ge 1\,,\\
\label{ZCC_L_k}
&&\partial_nL^{(k)}-\partial_kL^{(n)}+[L^{(k)},L^{(n)}]=0\,.
\eea
The common set of $2n$ PDEs for the $2n$ variables $b_j,c_j$, j$=1,\dots,n$ reads
\be
\label{common_PDEs}
\partial_kV_{\sharp,p}-\partial_nV_{\sharp,k+p-n}+\sum_{j=0}^{n-p}[V_{\sharp,n-j},V_{\sharp,k+p+j-n}]=0\,,~~p=1,\dots,n
\ee
where $\displaystyle V_{\sharp}^{(j)}=\sum_{m=0}^jV_{\sharp,m}\lda^{j-m}$ and $V_{\sharp}^{(j)}$ is either $V_{k}^{(j)}$ or $V_{n}^{(j)}$, $j=n,k$.
\end{proposition}
\prf
The discussion of the map $\Psi_k$ for any $k\ge 1$ in Section \ref{map_Psi} has established that it 
constrains the variable $b_j,c_j$ for $j>k$ and $a_j$ for all $j\ge 1$ in a specific manner. Hence, since $k>n$ here, it is not obvious a priori that 
\eqref{FNR_comp_k}-\eqref{ZCC_L_k} constrain the variables $b_j,c_j$ for $j=n+1,\dots,k$ in the same way as \eqref{FNR_comp_n}-\eqref{ZCC_L_n} do. The same remark 
goes for $a_j$, j$\ge 1$. That they do so is the key to obtain the complete equivalence of the two descriptions of \eqref{common_PDEs}.
Let us spell out \eqref{FNR_comp_n}-\eqref{ZCC_L_n},
\bea
\label{explicit_n}
\begin{cases}
\displaystyle\ell_0=\sigma_3\,,~~\partial_n\ell_p=\sum_{j=0}^n[\ell_j,\ell_{p+n-j}]\,,~~p\ge 1\,,\\
\displaystyle\partial_k\ell_{p}-\partial_n\ell_{k+p-n}+\sum_{j=0}^{n-p}[\ell_{n-j},\ell_{k+p+j-n}]=0\,,~~p=1,\dots,n\,,\\
\displaystyle\partial_n\ell_{p}=\sum_{j=0}^{n}[\ell_{n-j},\ell_{p+j}]\,,~~p=1,\dots,k-n\,.
\end{cases}
\eea
Of course, the third equation coming from the zero curvature is redundant here because of the first equation which is 
the FNR equation w.r.t. $t_n$. Upon solving the latter, \ie applying 
$\Psi_n$, we are therefore left with \eqref{common_PDEs} with $V_{n}^{(j)}$, $j=n,k$ with all their entries being polynomials in $b_j,c_j$, 
$j=1,\dots,n$ and {\it their derivatives w.r.t. $t_n$}. 

Consider now
\eqref{FNR_comp_k}-\eqref{ZCC_L_k}, equivalent to,
\bea
\label{explicit_k}
\begin{cases}
\displaystyle\ell_0=\sigma_3\,,~~\partial_k\ell_p=\sum_{j=0}^k[\ell_j,\ell_{p+k-j}]\,,~~p\ge 1\,,\\
\displaystyle\partial_k\ell_{p}-\partial_n\ell_{k+p-n}+\sum_{j=0}^{n-p}[\ell_{n-j},\ell_{k+p+j-n}]=0\,,~~p=1,\dots,n\,,\\
\displaystyle\partial_n\ell_{p}=\sum_{j=0}^{n}[\ell_{n-j},\ell_{p+j}]\,,~~p=1,\dots,k-n\,.
\end{cases}
\eea
We see that the first and third equations yield different information. For the $k$ relevant matrices
\be
\ell_j=\mato
a_j & b_j\\
c_j & -a_j
\matf\,,~~j=1,\dots,k
\ee
the first equation does not constrain $b_j,c_j$ $j=1,\dots,k$. It also gives expressions for $a_j$ 
in terms of $b_j,c_j$ $j=1,\dots,k$ and {\it their derivatives w.r.t. $t_k$}. So the situation is a priori very different from the 
first route. However, the third equation provides one part of the equations with respect to $t_n$ that we need. 
The required remaining equations w.r.t. $t_n$ are obtained by noting that the combination of the first equation for $p=1,\dots,n$ 
in \eqref{explicit_k} and the second equation yields
\be
\partial_n\ell_{p}=\sum_{j=0}^{n}[\ell_{n-j},\ell_{p+j}]\,,~~p=k-n+1,\dots,k\,.
\ee
Together with the third equation, this yields a total of $k$ equations with respect to $t_n$
\be
\label{truncated_n}
\partial_n\ell_{p}=\sum_{j=0}^{n}[\ell_{n-j},\ell_{p+j}]\,,~~p=1,\dots,k\,.
\ee
To complete the argument, we must show that this truncated system of equations constrains $b_j,c_j$, $j=n+1,\dots,k$ and $a_p$, $p=1,\dots,k$ 
in exactly the same way 
as the analogous non truncated system in \eqref{explicit_n} which defines the map $\Psi_n$. 

Consider first the $b,c$ variables. Since $\ell_0=\sigma_3$, for $p=1,\dots,k-n$, 
the truncated system determines recursively $b_j,c_j$ for $j=n+1,\dots,k$ in terms of $a_m^{(\ell)},b_m^{(\ell)},c_m^{(\ell)}$, $m\le n$, exactly as $\Psi_n$ does.

Consider now the $a$ variables. We must show that $a_p$, $p=1,\dots,n$ takes the same form as under $\Psi_n$. Upon examination of \eqref{truncated_n}, we see that $a_p$, $p=1,\dots,k$ determined by the truncated system
 does not depend on the entries of $\ell_m$ for $m>n$. Therefore, the truncated system can be embedded in the nontruncated system without altering the 
 expression for $a_p$, $p=1,\dots,k$. We can apply the argument in the non truncated system that yields the expressions 
of $a_p$ as polynomials in $a_m,b_m,c_m$, $m<p$ (meaning that the right-hand side of \eqref{FNR_comp_n} projected onto the diagonal is an exact $t_n$ derivative). 
When restricted to $a_p$, $p=1,\dots,k$, this yields the same expression as one would obtain via $\Psi_n$. 

We have now
shown that the two systems \eqref{explicit_n} and \eqref{explicit_k} yield the same generalized matrices $V_{k}^{(j)}$ and $V_{n}^{(j)}$, $j=n,k$.
In the first case, both are obtained by straightforward application of $\Psi_n$. In the second case, the missing equations with respect to $t_n$ are provided by the 
zero curvature condition itself, combined with the FNR equations with respect to $t_k$. 
\finprf

Combining Proposition \ref{equivalence_nk} with Proposition \ref{Hamiltonian_ZC} and Theorem \ref{equality_V} of the previous section, we can now state the main theorem of this section. 
\begin{theorem}
Let $n,k\ge 1$ be given. The set of PDEs obtained from $\Psi_n$ or $\Psi_k$, together with the corresponding zero curvature condition, has two equivalent Hamiltonian formulations
as
\be
\label{Ham_flow1}
\partial_k V_n^{(n)}=\{V_n^{(n)},H_n^{(k)}\}^{t_n}_R
\ee
and 
\be
\label{Ham_flow2}
\partial_n V_k^{(k)}=\{V_k^{(k)},H_k^{(n)}\}^{t_k}_R
\ee
where the two different Poisson brackets $\{~,~\}^{t_n}_R$ and $\{~,~\}^{t_k}_R$ are the brackets derived from $\Psi_n$ and $\Psi_k$ respectively. 
 $V_n^{(n)}$ (resp. $V_k^{(k)}$) satisfies the ultralocal Poisson algebra \eqref{Lax_matrix_bracket} with respect to $\{~,~\}^{t_n}_R$ (resp. $\{~,~\}^{t_k}_R$). 
 The Hamiltonian $H_n^{(k)}$ (resp. $H_k^{(n)}$) is constructed as in \eqref{pb_aux}-\eqref{def_Ham} (swapping the roles of $t_n$ and $t_k$ appropriately).
\end{theorem}
This theorem is the generalization to an arbitrary pair $t_n$, $t_k$ of the notion of dual Hamiltonian formulation that was introduced in \cite{ACDK}, 
based on explicit constructions for the pairs $n=1$, $k=2$ and $n=1$, $k=3$ with a special reduction ($b_1=c_1^*$) corresponding to the NLS equation and 
the complex modified KdV. In both cases, the variable $t_1$ was called $x$ and the corresponding Poisson bracket $\{~,~\}^{t_1}_R$ was called 
the equal-time bracket $\{~,~\}_S$. The variables $t_2$ and $t_3$ were taken in turn as the time $t$ and the corresponding Poisson bracket
$\{~,~\}^{t_k}_R$ was called the equal-space bracket $\{~,~\}^{(k)}_T$. The present approach puts all the times $t_n$ and $t_k$ on equal footing 
from the start.

\section{Examples}\label{examples}

\subsection{Recovering known results: the NLS case}

The NLS equation 
\be
iq_t+q_{xx}-2\epsilon|q|^2q=0\,~~\epsilon=\pm 1\,,
\ee
is well-known to follow from the FNR/AKNS scheme described previously by setting $n=1$, $k=2$. Solving the FNR equations w.r.t. $t_1$ and applying the 
zero curvature equation involving $t_2$ yields the following PDEs
\be
\label{NLS_not_red}
\begin{cases}
\partial_{t_2}b_1-\frac{1}{2}\partial^2_{t_1}b_1+b_1^2c_1=0\,,\\
\partial_{t_2}c_1+\frac{1}{2}\partial^2_{t_1}c_1-c_1^2b_1=0\,.
\end{cases}
\ee
Applying the reduction
\be
\label{reduction}
b_1=q\,,~~c_1=\epsilon q^*\,,
\ee
and setting $t_1=x$ and $t_2=2it$ gives NLS. There is no need to work only within this reduction so we will keep using $b_1$ and $c_1$ in the following, 
thus generalizing the results of \cite{CK,ACDK} which dealt only with the particular reduction \eqref{reduction}.

Here the two Lax matrices are, in the AKNS normalization, 
\be
V_1^{(1)}=\left(
\begin{array}{cc}
 \lambda  & b_1 \\
 c_1 & -\lambda  \\
\end{array}
\right)\,,~~V_1^{(2)}=\left(
\begin{array}{cc}
 \lambda ^2-\frac{1}{2} b_1 c_1 & \lambda  b_1+\frac{1}{2} \partial_{t_1}b_1 \\
 \lambda  c_1-\frac{1}{2} \partial_{t_1}c_1 & -\lambda ^2+\frac{1}{2} b_1 c_1 \\
\end{array}
\right)\,.
\ee
The Poisson bracket $\{~,~\}_R^{t_1}$ used for the Hamiltonian description of (non reduced) NLS is the famous one for which 
$b_1$ and $c_1$ are canonically conjugate variables. It yields the well-known ultralocal Poisson algebra \eqref{Lax_matrix_bracket} of Theorem \ref{theorem_Poisson_algebra}
 satisfied by $V_1^{(1)}$. In \cite{ACDK}, it was denoted $\{~,~\}_S$. The Hamiltonian $H_1^{(2)}$ obtained from \eqref{def_Ham} reads
\be
H_1^{(2)}=\frac{1}{16}\int_{-M}^M\left(b_1\partial_{t_1}^2c_1+b_1\partial_{t_1}^2c_1-2b_1^2c_1^2\right)\,dt_1
\ee
and the flow equation \eqref{Ham_flow1} boils down to
\bea
&&\partial_{t_2}b_1=\{b_1,H_1^{(2)}\}_R^{t_1}=\frac{1}{2}\partial_{t_1}^2b_1-b_1^2c_1\,,\\
&&\partial_{t_2}c_1=\{c_1,H_1^{(2)}\}_R^{t_1}=-\frac{1}{2}\partial_{t_1}^2c_1+c_1^2b_1
\eea
thus reproducing \eqref{NLS_not_red} as required. To compute the Poisson bracket of $b_1$, $c_1$ with the Hamiltonian $H_1^{(2)}$, we use
the fact that Poisson algebra \eqref{Lax_matrix_bracket} is equivalent to
\be
\{b_1(t_1),c_1(\tau_1)\}_R^{t_1}=4\delta(t_1-\tau_1)\,,
\ee 
and all other Poisson brackets being zero.

The dual Hamiltonian approach to (non reduced) NLS corresponds to swapping the role of $n$ and $k$. This yields the following 
two Lax matrices 
\be
V_2^{(1)}=\left(
\begin{array}{cc}
 \lambda  & b_1 \\
 c_1 & -\lambda  \\
\end{array}
\right)
\,,~~
V_2^{(2)}=\left(
\begin{array}{cc}
 \lambda ^2-\frac{1}{2} b_1 c_1 & \lambda  b_1+b_2 \\
 \lambda  c_1+c_2 & -\lambda ^2+\frac{1}{2} b_1 c_1 \\
\end{array}
\right)
\,.
\ee
As we expected, we work with four fields now in this dual picture. The zero curvature equation yields the following PDEs
\be
\begin{cases}
\partial_{t_1}b_1-2b_2=0\,,\\
\partial_{t_1}c_1+2c_2=0\,,\\
\partial_{t_2}b_1-\partial_{t_1}b_2+b_1^2c_1=0\,,\\
\partial_{t_2}c_1-\partial_{t_1}c_2-c_1^2b_1=0\,.
\end{cases}
\ee
We see in this example the procedure explained in full generality in Proposition \ref{equivalence_nk}. The first two equations produced by the 
zero curvature equation define the extra fields $b_2$, $c_2$ in terms of $b_1$, $c_1$. They are the FNR equations of the map $\Psi_1$ that 
are missing when we go over to the dual picture and apply the map $\Psi_2$. It is obvious that under the first two equations, the last two are 
the same as \eqref{NLS_not_red}. In \cite{ACDK}, we obtained a clear physical interpretation of the extra variables $b_2$, $c_2$ appearing in the 
dual formulation of NLS: they were the canonically conjugate momenta of $b_1=q$ and $c_1=q^*$ under the covariant Legendre transformation with respect to 
$t_1=x$. In the present work, they simply appear as additional coordinates in the reduced phase space whose construction from the 
infinite coadjoint orbit description of FNR we detailed in the previous section.

The dual Hamiltonian formulation of (non reduced) NLS is now obtained using the Poisson bracket $\{~,~\}_R^{t_2}$, the Hamiltonian 
\be
H_2^{(1)}=\frac{1}{2}\int_{-M}^M\left(b_2c_2+\frac{1}{4}(c_1\partial_{t_2}b_1-b_1\partial_{t_2}c_1+b_1^2c_1^2)\right)\,dt_2,,
\ee
and 
the flow equation \eqref{Ham_flow2} to obtain
\bea
&&\partial_{t_1}b_1=\{b_1,H_2^{(1)}\}_R^{t_2}=2b_2\,,\\
&&\partial_{t_1}c_1=\{c_1,H_2^{(1)}\}_R^{t_2}=-2c_2\,,\\
&&\partial_{t_1}b_2=\{b_2,H_2^{(1)}\}_R^{t_2}=\partial_{t_2}b_1+b_1^2c_1\,,\\
&&\partial_{t_1}c_2=\{c_2,H_2^{(1)}\}_R^{t_2}=\partial_{t_2}c_1-c_1^2b_1\,,
\eea
as required.
The Poisson bracket $\{~,~\}_R^{t_2}$ captures the ultralocal Poisson algebra \eqref{Lax_matrix_bracket} of Theorem \ref{theorem_Poisson_algebra} for 
$V_2^{(2)}$ now. In \cite{ACDK}, it was denoted $\{~,~\}_T^{(2)}$ and called the equal-space Poisson bracket. Here, it is equivalent to the 
following relations on the fields
\be
\{b_1(t_2),c_2(\tau_2)\}_R^{t_2}=4\delta(t_2-\tau_2) \,,~~\{c_1(t_2),b_2(\tau_2)\}_R^{t_2}=-4\delta(t_2-\tau_2)\,,
\ee
and all other Poisson brackets being zero. Under the special reduction \eqref{reduction}, the previous results reproduce those originally derived in \cite{CK,ACDK}.

The same analysis can be performed with $n=1$ and $k=3$. The results generalize those found in \cite{ACDK} for the 
complex modified KdV equation
\be
q_t+q_{xxx}-6\epsilon|q|^2q_x=0\,.
\ee
which is recovered by applying the reduction \eqref{reduction} and setting $t_1=x$ and $t_3=-4t$. The two Hamiltonians 
\be
H_1^{(3)}=\frac{1}{32}\int_{-M}^M\left(c_1\partial_{t_1}^3b_1-b_1\partial_{t_1}^3c_1+3b_1c_1(b_1\partial_{t_1}c_1-c_1\partial_{t_1}b_1) \right)\,dt_1
\ee
and
\be
H_3^{(1)}=\frac{1}{8}\int_{-M}^M\left(c_1\partial_{t_3}b_1-b_1\partial_{t_3}c_1+2b_1c_1(b_1c_2+b_2c_1)+4b_3c_2+4b_2c_3 \right)\,dt_3
\ee
produce the PDEs corresponding to the zero curvature for $(V_1^{(1)},V_1^{(3)})$ and $(V_3^{(1)},V_3^{(3)})$ using \eqref{Ham_flow1} and \eqref{Ham_flow2} 
respectively. Both systems boil down to
\bea
\label{cmKdV1}
&&\partial_{t_3}b_1-\frac{1}{4} \partial_{t_1}^3 b_1+\frac{3}{2} b_1c_1 \partial_{t_1}b_1 =0\,,\\
\label{cmKdV2}
&&\partial_{t_3}c_1-\frac{1}{4} \partial_{t_1}^3 c_1+\frac{3}{2} b_1c_1 \partial_{t_1}c_1 =0\,,
\eea
upon elimination of $b_2$, $c_2$, $b_3$, $c_3$ in the dual picture based on $(V_3^{(1)},V_3^{(3)})$. The Poisson brackets on the fields are
\be
\{b_1(t_1),c_1(\tau_1)\}=4\delta(t_1-\tau_1)\,,
\ee
as before, and
\bea
&&\{b_3(t_3),c_3(\tau_3)\}_R^{t_3}=-2b_1(t_3)c_1(t_3)\delta(t_3-\tau_3)\,,~~\{b_3(t_3),c_1(\tau_3)\}_R^{t_3}=4\delta(t_3-\tau_3)\,,\\
&&\{b_1(t_3),c_3(\tau_3)\}_R^{t_3}=4\delta(t_3-\tau_3)\,,~~\{b_2(t_3),c_2(\tau_3)\}_R^{t_3}=4\delta(t_3-\tau_3)\,,
\eea
in the dual picture.
Note that $b_3$ and $c_3$ are not canonical variables but rather obey a quadratic Poisson bracket 
relation. This is a rather general feature: one cannot expect in general that 
the variables $b_j$, $c_j$ inherited from the coadjoint orbit description of the FNR scheme are canonical variables for the associated 
Poisson bracket that we constructed. Finding such variables is a difficult task in general. However, in Section 7 of \cite{FNR}, the authors 
provide a set of variables (their equation $(40)$) that provide conjugate variables in their setting. It would be 
interesting to cast this aspect of their results into our classical $r$-matrix approach but that will not be done in this paper.

\subsection{A new example: Gerdjikov-Ivanov-type hierarchy}

We now apply our scheme to $n=2$ and $k=4$ . This gives rise to equations that generalize those studied in \cite{GI} 
which are related to a wealth of NLS-type equations: derivative NLS equation of the Kaup-Newell hierarchy \cite{KN}, the Chen-Lee-Liu equation \cite{CLL}, the WKI equation \cite{WKI} 
via gauge transformations \cite{K}. This shows that the Gerdjikov-Ivanov equation can be embedded in a hierarchy whose classical $r$-matrix structure 
is of the rational type used all along this paper, a feature that is new to the best of our knowledge. Of course, it is known that some 
reductions (like the ones we will apply below) and gauge transformations do not preserve the $r$-matrix structure. This example also illustrates a point 
that we mention in the next section and that deserves a more careful analysis, going beyond the scope of the present paper: thanks to our dual approach, one 
can ``travel'' in a multidimensional lattice of Lax pairs and make connections between hierarchies that are traditionally thought of as distinct. One ends up 
with multiple multiHamiltonians hierarchies.

The Lax pair corresponding to $t_2$ and $t_4$ reads
\bea
&&V_2^{(2)}=\left(
\begin{array}{cc}
 \lambda ^2-\frac{1}{2} b_1 c_1 & \lambda  b_1+b_2 \\
 \lambda  c_1+c_2 & -\lambda ^2+\frac{1}{2} b_1 c_1 \\
\end{array}
\right)\,,\\
&&V_2^{(4)}=\left(
\begin{array}{cc}
\alpha(\lda) & \beta(\lda)\\
\gamma(\lda) & -\alpha(\lda)
\end{array}
\right)\,,
\eea
with 
\bea
&&\alpha(\lda)= \lambda ^4-\frac{1}{2} b_1 c_1 \lambda ^2-\frac{1}{2}\left( b_2 c_1+ b_1 c_2\right) \lambda +\frac{1}{8} \left(2 b_1 \partial_{t_2}c_1-2 \partial_{t_2}b_1 c_1-4 b_2 c_2-b_1^2 c_1^2\right)\,,\\
&&\beta(\lda)=b_1 \lambda ^3+b_2 \lambda ^2+\frac{1}{2} \partial_{t_2}b_1 \lambda +\frac{1}{2} \left(\partial_{t_2}b_2-c_2 b_1^2-b_2 c_1 b_1\right)\,,\\
&&\gamma(\lda)=c_1 \lambda ^3+c_2 \lambda ^2-\frac{1}{2} \partial_{t_2}c_1 \lambda -\frac{1}{2} \left(\partial_{t_2}c_2+b_2 c_1^2+b_1 c_2 c_1\right)\,.
\eea
The corresponding PDEs read
\bea
\label{GI_system}
&&\partial_{t_4}b_1-\frac{1}{2}\partial_{t_2}^2b_1+b_2^2c_1+2b_2c_2b_1-\frac{1}{2}b_1^2\partial_{t_2}c_1+\frac{1}{4}c_1^3b_1^2=0\,,\\
&&\partial_{t_4}c_1+\frac{1}{2}\partial_{t_2}^2c_1-c_2^2b_1-2c_2b_2c_1-\frac{1}{2}c_1^2\partial_{t_2}b_1-\frac{1}{4}b_1^3c_1^2=0\,,\\
&&\partial_{t_4}b_2-\frac{1}{2}\partial_{t_2}^2b_2+b_2^2c_2+(b_1c_2+b_2c_1)\partial_{t_2}b_1+\frac{1}{2}b_1^2\partial_{t_2}c_2+\frac{3}{2}b_1^2c_1^2b_2+\frac{1}{2}b_1^3c_1c_2=0\,,\\
&&\partial_{t_4}c_2+\frac{1}{2}\partial_{t_2}^2c_2-c_2^2b_2+(b_1c_2+b_2c_1)\partial_{t_2}c_1+\frac{1}{2}c_1^2\partial_{t_2}b_2-\frac{3}{2}b_1^2c_1^2c_2-\frac{1}{2}c_1^3b_1b_2=0\,.
\eea
In the special case $b_2=c_2=0$, this reduces to
\bea
&&\partial_{t_4}b_1-\frac{1}{2}\partial_{t_2}^2b_1-\frac{1}{2}b_1^2\partial_{t_2}c_1+\frac{1}{4}b_1^3c_1^2=0\,,\\
&&\partial_{t_4}c_1+\frac{1}{2}\partial_{t_2}^2c_1-\frac{1}{2}c_1^2\partial_{t_2}b_1-\frac{1}{4}c_1^3b_1^2=0\,.
\eea
The further reduction $b_1=q$, $c_1=i\epsilon q^*$, $\epsilon=\pm 1$ and the choice $t_2=x$, $t_4=2it$ yield the Gerdjikov-Ivanov equation
\be
iq_t+q_{xx}+i\epsilon q^2q^*_x+\frac{1}{2}|q|^4q=0\,.
\ee
As before it can be checked that the system of PDEs \eqref{GI_system} is of Hamiltonian type with Hamiltonian
\bea
H_2^{(4)}&=&
\frac{1}{16}\int_{-M}^M \left(b_2 \partial_{t_2}^2c_1+ c_2\partial_{t_2}^2b_1  + c_1\partial_{t_2}^2b_2+b_1 \partial_{t_2}^2c_2\right)\,dt_2\\
&&+\frac{1}{32}\int_{-M}^M \left( b_1c_1(b_2\partial_{t_2}c_1-c_2\partial_{t_2}b_1) +
3 b_1^2c_2 \partial_{t_2}c_1-3b_2c_1^2 \partial_{t_2}b_1\right)\,dt_2\\
&&+\frac{1}{32}\int_{-M}^M \left(-2 b_1^3 c_1^2 c_2- 2 b_1^2b_2 c_1^3 -8 b_1b_2c_2^2-8 c_1 c_2b_2^2 \right)\,dt_2
\eea
and Poisson brackets between the fields as 
\be
\{b_1(t_2),c_2(\tau_2)\}_R^{t_2}=4\delta(t_2-\tau_2) \,,~~\{c_1(t_2),b_2(\tau_2)\}_R^{t_2}=-4\delta(t_2-\tau_2)\,,
\ee
all other Poisson brackets being zero. We see that these are the same Poisson brackets as in the NLS case in the dual formulation. This is of course 
consistent with the fact that the latter is also based on $V_2^{(2)}$. In fact, NLS in the dual formulation is nothing but the first level in the same hierarchy
as the present one: one simply picks the $t_1$ flow with associated Lax matrix $V_2^{(1)})$ instead of the $t_4$ flow used here. Roughly speaking, 
we can ``travel'' from the standard NLS hierarchy based on $V_1^{(1)}$ to the present different hierarchy based on $V_2^{(2)}$ via the dual formulation of NLS which 
precisely shifts the emphasis from $V_1^{(1)}$ to $V_2^{(2)}$. Our general results show that the $r$-matrix structure follows in the process.

The details of the dual formulation of the system \eqref{GI_system} are not very illuminating in themselves and only confirm the consistency of our general picture. 
The PDEs \eqref{GI_system} are recovered from the zero curvature based on the Lax pair $(V_4^{(2)},V_4^{(4)})$ and from the Hamiltonian flow \eqref{Ham_flow2} 
with respect to $\{~,~\}_R^{t_4}$ using the Hamiltonian $H_4^{(2)}$.

\section{Discussion and conclusion}

\subsection{Outlook}

The puzzling observation that the two Lax matrices involved in a given zero curvature condition each satisfy the same ultralocal 
Poisson algebra (albeit with respect to very different and not compatible Poisson brackets) was the main motivation for the present paper. In the 
process of understanding this, we have now uncovered the general features of the diagram shown in the introduction. 
The appearance of the same $r$-matrix structure in the Poisson brackets of the two Lax matrices is traced back to the R operator formulation of the FNR 
scheme. In addition, we made no use of the Lagrangian formalism here and the associated cumbersome covariant Legendre transformations and Dirac procedure, but have only used algebraic tools.

Of immediate attention is now the question of generalizing the results of this paper to other types of classical $r$-matrix contexts: trigonometric $r$-matrix and non-ultralocal $r$-matrix. 
It is not clear how our approach can be extended to those cases since the specific pole structure of the $r$-matrix seems to play a crucial role in several demonstrations. However,
 at least the trigonometric extensions have in many cases suitable interpretations in terms of coadjoint orbit constructions, which gives us some hope regarding a partial extension to this case. The elliptic 
case seems much more tricky.

\subsection{Covariance and connection with results on dressing transformations}

Combining the results of the present paper (purely algebraic) with the previous results of \cite{ACDK} (purely field theoretic), we are naturally led to suggest that the classical $r$-matrix has 
a fundamental covariant nature that has not been investigated so far. Since its discovery, it has always been tied to a traditional Hamiltonian approach which singles out one 
time variable. Our paper 
represents a first step at elucidating the covariant nature of the classical $r$ matrix. To our knowledge, the most advanced theory of integrable systems from a covariant field theory point of view has 
been achieved in Chapter $19$ of \cite{Dickey_book}. However, the classical $r$-matrix is totally absent there. In our opinion, these two important topics deserve to be fused into a consistent integrable 
covariant field theory. 

A possible direction of investigation towards understanding this $r$-matrix ``universality'' uncovered in this paper 
may be suggested by general features of classical integrable systems, 
namely the structures of the associated dressing transformations.
Let us indeed summarize precisely our results and compare them with some known facts on dressing transformations of integrable PDEs.

We have established that for any choice of a continuous variable $x \equiv t_n$ amongst the hierarchy of time variables in the FNR formalism
the linear Poisson structure of the constrained Lax matrix ($x$-component of the Lax connection) and therefore the subsequent quadratic Poisson structure of 
the associated monodromy matrix 
were described by the $r$ matrix $\displaystyle r_{12} = \frac{\Pi_{sl(2,C)}}{\lambda - \mu}$. It is in particular true of the two Lax matrices in a pair
of dual Lax representations such as were discussed in Section $3.3$, yielding two zero-curvature representations of the same nonlinear integrable PDE. 

We argue that this universality of the $r$ matrix structure in a dual pair could have been proposed ab initio from arguments based on the general properties of the group of dressing
transformations acting on the set of solutions to the PDE (possibly endowed with a manifold structure if a moduli space structure can be defined). 

The relevant general results on dressing transformations are formulated as follows: 

Assuming that
some integrable  PDE is described in a AKNS-ZS framework by a zero-curvature condition of a connection $ \{ d/dx - L(\lambda), d/dt - M(\lambda) \}$ depending on one complex spectral
parameter $\lambda$, it was proved that the dressing group was:

1. Isomorphic as a group to the direct product of groups of germs of Lie algebra-valued analytic functions in $\lambda $ at each (simple) pole in $\lambda$ of the Lax pair \cite{AB}.

2. Endowed with a Lie-Poisson structure directly obtained from the Lie-Poisson structure of the monodromy matrix identified as a generator of the dressing group \cite{BBT,STS2}, and consistent
with the algebraic group structure.

3. In addition, the pole structure of the classical Lax matrix is determined by the precise structure of the Adler-Kostant-Symes split, defining intrinsically the integrability structure, and yielding directly the $R$-operator, hence the $r$ matrix.

Given that the dressing group acts on the (moduli) space of the solutions to the PDE, it is expected to be only sensitive (as a group) to the Lagrangian aspect and not to any Hamiltonian 
or zero-curvature representation of the PDE. Hence both its own algebraic structure and associated Lie-Poisson structure are a priori intrinsic.
At the same time the algebraic structure and Lie-Poisson structure are technically obtained \cite{AB} from the poles of the Lax matrix which themselves determine 
the poles of the $r$-matrix by the duality procedure between $r$-matrix and $R$-operator, as discussed in Point 3. Finally, the Lie-Poisson structure of the dressing group allows one to turn the action of the dressing group on the space of solutions into a Poisson action \cite{STS2}, implying that the Poisson structure (whatever it is) underlying the integrable PDE must exhibit some compatibility property with respect to the Lie-Poisson structure of the dressing group.

This strongly suggests that whichever choice of Hamiltonian structure and associated Lax representation is picked for a PDE (within some limits to be investigated more precisely, in particular regarding the Lie algebra structure involved in the construction), we should expect the $r$-matrix structure to always be consistent with the fundamental algebraic features, described in Points 1 and 2, of the dressing transformation group. Hence it should be independent of this particular choice; as is indeed the case for the ``dual'' representations. 

To summarize this informal discussion, it is not surprising {\it a posteriori} to observe the same $r$-matrix structure for the two Lax matrices involved in the description of a given integrable PDE, if one relates it 
to the dressing transformation group of that PDE since the latter is not sensitive to the distinction between an $x$ or $t$ variable.

\subsection{Multiple hierarchies}

Having established the pattern to get any ``space'' Lax operator from a chosen time-evolution in a coadjoint orbit formulation, we can pose 
the problem of connecting the construction of two non-compatible Poisson brackets associated with each term of a Lax pair, to the well-established theory of 
hierarchies of Poisson structures a la Magri attached to a given Lax matrix endowed with its Poisson structure. Let us describe the issue more precisely.

Start from an initial dual Lax pair compactly denoted $(L,M)$. From the monodromy of each of our space Lax operator $L$ or $M$, together with its Poisson structure and $r$-matrix, we generate a hierarchy of mutually commuting time evolutions. They have been shown here to be also obtained by applying the relevant constraint-extension map $\Psi$ 
to the corresponding operator in the coadjoint orbit framework. 

Consider now a pair made of the initial space Lax operator $L$ and one of its time Lax operator $M^{(k)}$ in the hierarchy. Given our results the duality property still holds, 
hence there exists a Poisson structure under which $M^{(k)}$ becomes a ``space'' operator and $L$ a ``time'' operator. We may then use the monodromy of $M^{(k)}$ to compute a hierarchy of new time operators now denoted
$L^{(k,s)}$ . Proceeding
similarly with the initial ``time'' operator $M$ of the dual pair now taken as ``space'' operator, we can build from its monodromy a hierarchy of new operators $L^{(s')}$ , any of which may then be taken by duality as ``space'' operator to generate the operators $M^{(s',k')}$. Repeating the process in this way, we should  be able to build a multilabel structure of Lax-type operators.

The beginning of this construction was undertaken in the previous paper \cite{ACDK}. A key question is the following: is this multilabel structure a ``flat'' lattice? In other words, can one 
 identify operators $L^{(k,s)}$ and $M^{(s',k')}$ for instance (provided $s=s', k=k'$ or maybe some weaker condition)? More generally, does the previous process close after a finite number of steps? 
This already seemed not to be the case in the simplest situations. In fact as we 
have seen the identification of operators obtained by monodromy construction versus operators obtained by constraint-extensions is valid on-shell as is clear from the formulation of Proposition \ref{equivalence_nk}. Only at the level of equations can we establish the equivalence. This point thus clearly requires a very careful examination, which is left for future work.

\section*{Ackowledgements}

We thank L-C Li for his useful comments on a draft of this paper.  
V.C. acknowledges the hospitality of the University of Cergy-Pontoise and l'Institut des Etudes Avanc\'ees for supporting his visit at the LPTM where most of this research was carried out.
V.C. is indebted to A. Fordy for stimulating discussions and for drawing his attention to the theory of stationary manifolds and the references \cite{FH1,FH2}.
We are also indebted to M. Semenov-Tian-Shansky for reading the first version of our preprint and pointing out references \cite{R,KR} that helped us put our 
results into perspective.

\end{document}